\documentclass[12pt,a4paper,twoside]{article}

\usepackage[dvips]{graphics}
\usepackage{cite}
\usepackage{color}

\usepackage{bigstrut}
\usepackage{amssymb}
\usepackage{rotating}

\newcommand{\epem}     {\ensuremath{\mathrm{e^+e^-}}}
\newcommand{\roots}    {\ensuremath{\sqrt{s}}}

\newcommand{\znull}    {\ensuremath{\mathrm{Z^0}}}
\newcommand{\mz}       {\ensuremath{m_{\znull}}}
\newcommand{\as}       {\ensuremath{\alpha_{\mathrm{S}}}}
\newcommand{\ash}      {\ensuremath{\hat{\alpha}_{\mathrm{S}}}}
\newcommand{\asmz}     {\ensuremath{\as(\mz)}}

\newcommand{\stat}     {\ensuremath{\mathrm{(stat.)}}}

\newcommand{\expt}     {\ensuremath{\mathrm{(exp.)}}}
\newcommand{\had}      {\ensuremath{\mathrm{(had.)}}}
\newcommand{\theo}     {\ensuremath{\mathrm{(theo.)}}}
\newcommand{\xmu}      {\ensuremath{x_{\mu}}}
\newcommand{\bt}       {\ensuremath{B_{\mathrm{T}}}}
\newcommand{\bw}       {\ensuremath{B_{\mathrm{W}}}}
\newcommand{\mh}       {\ensuremath{M_{\mathrm{H}}}}

\newcommand{\thr}      {\ensuremath{1-T}}
\newcommand{\cp}       {\ensuremath{C}}

\newcommand{\ytwothree}{\ensuremath{y_{23}^\mathrm{D}}}

\newcommand{\oaaa}     {\ensuremath{\mathcal{O}(\as^3)}}
\newcommand{\chisq}  {\ensuremath{\chi^2}}
\newcommand{\chisqd} {\ensuremath{\chi^2/\mathrm{d.o.f.}}}

\newcommand{\Opal}{\mbox{\rm OPAL}}

\newcommand{\muR} {\ensuremath{\mu_R}}
\newcommand{\rs}    {\ensuremath{\sqrt{s}}}
\newcommand{\xL}{\ensuremath{x_{_L}}}
\newcommand{\betweenline}{\\ \noalign{\smallskip}\hline\noalign{\smallskip}}
\newcommand{\ww}     {\ensuremath{\mathrm{W^+W^-}}}
\newcommand{\qqbar}  {\ensuremath{\mathrm{q\overline{q}}}}

\newcommand{\Herwig}{\mbox{HERWIG}}
\newcommand{\Pythia}{\mbox{PYTHIA}}
\newcommand{\Jetset}{\mbox{JETSET}}

\newcommand{\Koralw}{\mbox{KORALW}}
\newcommand{\kkff}{\mbox{$\mathcal{KK}$2f}}
\newcommand{\Ariadne}{\mbox{ARIADNE}}
\newcommand{\grcff}{\mbox{grc4f}}

\newcommand{\resultnnloA}{\ensuremath{ 0.1201\pm 0.0008\stat\pm 0.0013\expt}}
\newcommand{\resultnnloB}{\ensuremath{\pm 0.0010\had\pm 0.0024\theo      }}
\newcommand{\resulttotnnlo}{\ensuremath{\asmz= 0.1201\pm 0.0031}}
\newcommand{\resultnnlonllaA}{\ensuremath{ 0.1189\pm 0.0008\stat\pm 0.0016\expt}}
\newcommand{\resultnnlonllaB}{\ensuremath{\pm 0.0010\had\pm 0.0036\theo      }}
\newcommand{\resulttotnnlonlla}{\ensuremath{\asmz= 0.1189\pm 0.0041}}
\begin{document}
%  Title Page:
\begin{titlepage}
\begin{center}{\large   EUROPEAN ORGANIZATION FOR NUCLEAR RESEARCH
}\end{center}\bigskip
\begin{center}
{\huge\bf
Determination of \boldmath{\sc\as} using OPAL hadronic event shapes
at $\rs = 91-209$~GeV and resummed NNLO calculations}

\end{center}
\bigskip\bigskip
\begin{center}
{\LARGE The OPAL Collaboration}
\end{center}
\bigskip\bigskip\bigskip
\begin{center}{\large  Abstract}\end{center}
Hadronic event shape distributions from \epem\ annihilation
measured by the OPAL experiment at centre-of-mass energies between
91~GeV and 209~GeV are used to determine the strong coupling \as.  
  The results are based on QCD predictions complete to the next-to-next-to-leading 
  order (NNLO), and on NNLO calculations matched to the resummed 
  next-to-leading-log-approximation terms (NNLO+NLLA).
The combined NNLO result from all variables and centre-of-mass energies is
\begin{displaymath} 
  \asmz=\resultnnloA\resultnnloB.
\end{displaymath} 
while the combined NNLO+NLLA result is
\begin{displaymath}
  \asmz=\resultnnlonllaA\resultnnlonllaB.
\end{displaymath}
The completeness of the NNLO and NNLO+NLLA results with respect to
missing higher order contributions, 
studied by varying the renormalization scale,
is improved
compared to 
previous results based on NLO or NLO+NLLA predictions only.
The observed
energy dependence of \as\ agrees with the QCD prediction of asymptotic
freedom and excludes the absence of running.
\bigskip\bigskip
\begin{center}
{\large 
(To be submitted to European Physical Journal C) }
\end{center}
%
%  Authors and time stamp:
\begin{center}

\vspace*{0.5 cm}

{\large final draft} \\
\today
\bigskip

\end{center}
\end{titlepage}

\begin{center}{\Large        The OPAL Collaboration
}\end{center}\bigskip
\begin{center}{
G.\thinspace Abbiendi$^{  2}$,
C.\thinspace Ainsley$^{  5, u}$,
P.F.\thinspace {\AA}kesson$^{  7}$,
G.\thinspace Alexander$^{ 21}$,
G.\thinspace Anagnostou$^{  1}$,
K.J.\thinspace Anderson$^{  8}$,
S.\thinspace Asai$^{ 22}$,
D.\thinspace Axen$^{ 26}$,
I.\thinspace Bailey$^{ 25, g}$,
E.\thinspace Barberio$^{  7,   o}$
% ...
%end authorlist PLEASE DO NOT DELETE THIS COMMENT
}\end{center}\bigskip
\bigskip
%begin institutes
$^{  1}$School of Physics and Astronomy, University of Birmingham,
Birmingham B15 2TT, UK
\newline
$^{  2}$Dipartimento di Fisica dell' Universit\`a di Bologna and INFN,
I-40126 Bologna, Italy
\newline
% ...
$^{  5}$Cavendish Laboratory, Cambridge CB3 0HE, UK
\newline
% ...
$^{  7}$CERN, European Organisation for Nuclear Research,
CH-1211 Geneva 23, Switzerland
\newline
$^{  8}$Enrico Fermi Institute and Department of Physics, University of Chicago, Chicago IL 60637, USA
\newline
% ...
$^{ 21}$Department of Physics and Astronomy, Tel Aviv University,
Tel Aviv 69978, Israel
\newline
$^{ 22}$International Centre for Elementary Particle Physics and
Department of Physics, University of Tokyo, Tokyo 113-0033, and
Kobe University, Kobe 657-8501, Japan
\newline
% ...
$^{ 25}$University of Victoria, Department of Physics, P O Box 3055,
Victoria BC V8W 3P6, Canada
\newline
$^{ 26}$University of British Columbia, Department of Physics,
Vancouver BC V6T 1Z1, Canada
\newline
% ...
$^{  g}$ now at Lancaster University
\newline
% ...
$^{  o}$ now at The University of Melbourne, Victoria, Australia
\newline
% ...
$^{  u}$ now at University of Pennsylvania, Philadelphia, Pennsylvania, USA
\newline
% ...
%end notes

\newpage

%
%
%  The main text:
%

\section{Introduction}

Events originating from \epem\ annihilation into quark-antiquark pairs
allow 
precision tests~\cite{OHab,STKrev,dissertori03} of 
the theory of the strong interaction,
Quantum Chromodynamics (QCD)~\cite{FritzschGellMann,GrossWilczek1,GrossWilczek2,Politzer}.
Comparison of observables like jet production rates or event shape variables
with theoretical predictions allows %%to determine 
the crucial free parameter of QCD -- the strong coupling \as\ -- to be determined.
The determination of \as\ from many different observables
provides an important consistency test of QCD.
Recently, significant progress in the
theoretical calculation of event shape observables has been made. %% and
Next-to-next-to-leading order (NNLO) calculations are now
available~\cite{NNLOESs,Weinzierl} as well as NNLO calculations matched with resummed
terms in the next-to-leading-logarithmic-approximation 
(NLLA)~\cite{NNLONLLAES}.  

Measurements of \as\ at centre-of-mass-system (c.m.) energies between 
$\rs=91~$GeV and  $206~$GeV, using NNLO predictions and ALEPH data, were 
presented in~\cite{asNNLO}. This was followed by measurements between 
$\roots=14$~GeV and $\roots=44$~GeV~\cite{jadeNNLO} and between 
$\roots=91$~GeV and $\roots=206$~GeV~\cite{alephNNLONLLA} based on comparing 
JADE 
or ALEPH
data with NNLO plus matched NLLA calculations.  
The strong coupling has also been determined at NNLO from the three-jet rate
at LEP~\cite{threeJetNNLO}.
In the present study, the revised NNLO calculations described
  in~\cite{NNLOESs,Weinzierl} are used to determine \as\ at NNLO and NNLO+NLLA for 13
  different energy values between 91 and 209~GeV.   
  The data sample is that of the OPAL Collaboration at LEP.  Hadronization
  corrections are treated in the manner described in~\cite{STKrev,jadeNNLO,asNNLO,OPALPR404}. The
  study is based on the same measurements of event shape variables, the same fit
  ranges (except for \ytwothree\ as discussed below), and the same Monte Carlo models for detector 
  and hadronization corrections as those in~\cite{OPALPR404}.

The structure of the paper is as follows.
In Sect.~\ref{sec_detector} we give an overview of the \Opal\
detector, and in Sect.~\ref{sec_datamc} we summarize the data and
Monte Carlo samples used.  The theoretical background to the work is 
outlined in Sect.~\ref{sec_theo}.  The experimental analysis 
techniques are explained in 
Sect.~\ref{sec_anal}, and the measurements
are compared with theory in Sect.~\ref{sec_results}.

\section{The OPAL Detector\label{sec_detector}}
The OPAL experiment operated from 1989 to 2000 at the LEP 
\epem\ collider at CERN. 
The OPAL detector is described in
detail in~\cite{opaltechnicalpaper1,opaltechnicalpaper2,opaltechnicalpaper3}.
  This analysis mostly makes use of the measurements of energy deposited
  in the electromagnetic calorimeter and of charged particle momenta in
  the tracking chambers.

All tracking systems were located inside a solenoidal magnet, which
provided a uniform axial magnetic field of 
0.435~T along the beam
axis.
The main tracking detector was the central jet chamber. This device was
approximately 4~m long and had an outer radius of about 1.85~m. 
The magnet was surrounded by a lead glass electromagnetic
calorimeter and a sampling had\-ron calorimeter.  
The electromagnetic calorimeter 
consisted of 
11704
lead glass blocks, covering 98\% of the solid angle.
Outside
the hadron calo\-ri\-me\-ter, the detector was surrounded by a system of muon
chambers. 
 
\section{Data and Monte Carlo Samples\label{sec_datamc}}
The analysis is based on 
the same data, Monte Carlo samples, and event selection, 
as those employed in a previous study~\cite{OPALPR404}. The data were collected
at c.m.\ energies between 91.0~GeV and 208.9~GeV. 
  The data at 91~GeV, on the \znull\ peak, were collected during calibration
  runs between 1996 and 2000 and have the same conditions of detector
  configuration, performance and software reconstruction as the higher 
  energy data.
The data are grouped into samples with similar c.m.\ energies, as
indicated in Tab.~\ref{TableNumberEvents}.  Besides the energy range, mean energy and
integrated luminosity, Tab.~\ref{TableNumberEvents} lists the year(s) of collection and
the number of selected hadronic annihilation events for each sample.

Samples of Monte Carlo-simulated events were used to
correct the data for experimental acceptance, efficiency and backgrounds.
The $\epem\rightarrow\qqbar$ process was simulated at $\sqrt{s}=91.2$~GeV using
\Jetset~7.4~\cite{jetset3}, and at higher energies
using \kkff~4.01 or \kkff~4.13~\cite{kk2f1,kk2f2} with 
hadronization performed using \Pythia~6.150 or 
PY-\linebreak THIA~6.158~\cite{jetset3}.
Corresponding samples of HERWIG~6.2\cite{herwig1,herwig6} or \kkff\ events 
with \Herwig~6.2 hadronization were used for systematic checks.  
Four-fermion background processes were simulated using \grcff~2.1~\cite{grc4f}, \Koralw~1.42~\cite{koralw} 
with \grcff~\cite{grc4f} matrix elements, with hadronization 
performed using 
\Pythia.  The above samples, generated at each energy point studied, 
were processed through a full simulation of the \Opal\ 
detector~\cite{gopal} and reconstructed in the same manner as the data.
In addition, 
when correcting for the effects of hadronization, large samples
of generator-level Monte Carlo events were employed, using the
parton shower 
models \Pythia~6.158, \Herwig~6.2 and 
\Ariadne~4.11\cite{ariadne3}.

Each of these
models used for the description of hadronization and detector response 
contains a number of tunable parameters.
These parameters were adjusted to describe previously published \Opal\ data at 
$\sqrt{s}\sim91$~GeV as discussed in~\cite{OPALPR141} for 
\Pythia/ \Jetset\ and in~\cite{OPALPR379} for \Herwig\ and \Ariadne.

\section{Theoretical Background\label{sec_theo}}
\subsection{Event Shape Distributions}
The properties of hadronic events can be described by event shape
observables.  The event shape observables used for this analysis are
thrust (\thr) ~\cite{thrust1,thrust2}, heavy jet mass
(\mh)~\cite{def_mh1}, wide and total jet broadening (\bw\ and
\bt)~\cite{nllabtbw}, C-Parameter (\cp)~\cite{def_c1,def_c2,ert}
and the transition value between 2 and 3 jet
configurations defined using the Durham jet
algorithm (\ytwothree)~\cite{durham}.  

In the following, the symbol $y$ is used to refer to any of the variables  
\thr, \mh, \bt, \bw, \cp\ or \ytwothree.
  Event shape variables characterize the main features of the
  distribution of momentum in an event. Larger values of $y$ correspond to the multi-jet region
  dominated by the radiation of hard  
  gluons.  Smaller values of $y$ correspond to the two-jet region 
  with only soft and collinear radiation.

\subsection{QCD Calculations}
\label{qcdprediction}
QCD predictions for the distribution of event shape observables
in \epem\ annihilations are now available to \oaaa\ (NNLO)~\cite{NNLOESs,Weinzierl}.
In the case where the renormalization scale \muR\ equals the physical scale 
$Q$ identified with \roots, the predictions are:
\begin{equation}
  \left.\frac{1}{\sigma}\frac{{\rm d} \sigma}{{\rm d}y}\right|_{\oaaa} 
    = \frac{{\rm d}A}{{\rm d}y}\ash + \frac{{\rm d}B}{{\rm d}y}\ash^2 + \frac{{\rm d}C}{{\rm d}y}\ash^3
\label{NNLOcalc}        
\end{equation}
with $\ash=\as(\muR)/(2\pi)$.  
The coefficient distributions for the leading order (LO)
${\rm d}A/{\rm d}y$, next-to-leading order (NLO) ${\rm d}B/{\rm d}y$ and NNLO ${\rm d}C/{\rm d}y$ 
terms were provided to us by the authors of~\cite{NNLOESs}.
The
normalization to the total hadronic cross section and the terms
generated by variation of the renormalization scale parameter
$\xmu=\muR/Q$ are implemented according to the prescription of~\cite{NNLOESs}.

In the two-jet (low $y$) region, the effect of soft and 
collinear emissions introduces large logarithmic 
effects depending on $L=\log(1/y)$. 
For 
the $(1-T)$, \mh, \bt, \bw, $C$ and
\ytwothree\ distributions, 
the leading and next-to-leading logarithmic terms can be resummed up to infinite
order in perturbation theory~\cite{resummation}.
This is referred to as the next-to-leading-logarithmic approximation (NLLA).
The most complete calculations of event shape observables are obtained from combining (matching) the
\oaaa\ and NLLA calculations, taking care not to double count 
terms that are in common between them.

The matching is not unique. 
In the so-called ln$R$
matching scheme, the NNLO+NLLA expression for the logarithm 
of the 
cumulative distribution
$R(y)=\int_0^y {\rm d}y'(1/\sigma) {\rm d}\sigma/ {\rm d}y'$
is~\cite{NNLONLLAES}
\begin{eqnarray}\label{logRmatching}
&\ln&\left(R\left(y,\ash\right)\right)=L\,g_{1}\left(\ash L\right)\,+\,g_{2}\left(\ash L\right)\\
&+&\,\ash\left({A}\left(y\right)-G_{11}L-G_{12}L^{2}\right){}\nonumber\\
&+&\,\ash^{2}\left({B}\left(y\right)-\frac{1}{2}{A}^{2}\left(y\right)-G_{22}L^{2}-G_{23}L^{3}\right){}\nonumber\\
&+&\,\ash^{3}\left({C}\left(y\right)-{A}\left(y\right){B}\left(y\right)+\frac{1}{3}{A}^{3}\left(y\right)-G_{33}L^{3}-G_{34}L^{4}\right).\nonumber
\end{eqnarray}
The functions $g_1$ and $g_2$ represent the resummed leading and next-to-leading logarithmic terms   
while the $G_{ij}$ are matching coefficients. 
The coefficient functions $A$, $B$ and $C$ are related to the differential coefficients in~(\ref{NNLOcalc})
by integration, e.g. $A(y)=\int_0^y {\rm d}y'\frac{{\rm d}A(y')}{{\rm d}y'}$.
Since
 \as\ in the NNLO and NLLA terms are assumed to be the same, 
there is only one renormalization scale.
The NLLA terms introduce a further arbitrariness in the choice of a logarithmic rescaling
variable \xL, see~\cite{OPALPR404}.

The theoretical calculations provide distributions at the level of
quarks and gluons, the so-called parton-level.  Monte Carlo-based
distributions calculated using the final-state partons after
termination of the parton showering in the models are also said to be
at the parton-level.  
In contrast, the data are corrected to the hadron-level,
i.e.\ they correspond to the distributions of the stable particles
in the event as explained in Sect.~\ref{detcorr}.
To compare the QCD predictions with measured 
event shape distributions these predictions are corrected 
from the parton to the hadron level.
The corrections are based on large samples of typically $10^7$ events generated with
the parton-shower Monte Carlo programs PYTHIA (used by default),
HERWIG and ARIADNE 
(used to evaluate systematic uncertainties).
The corrections are defined by the ratio of the Monte Carlo distributions
at the hadron and parton levels and are applied as multiplicative 
corrections to the theoretical predictions.

\begin{figure*}[htb!]
\begin{center}
\includegraphics[width=1.\columnwidth,clip]{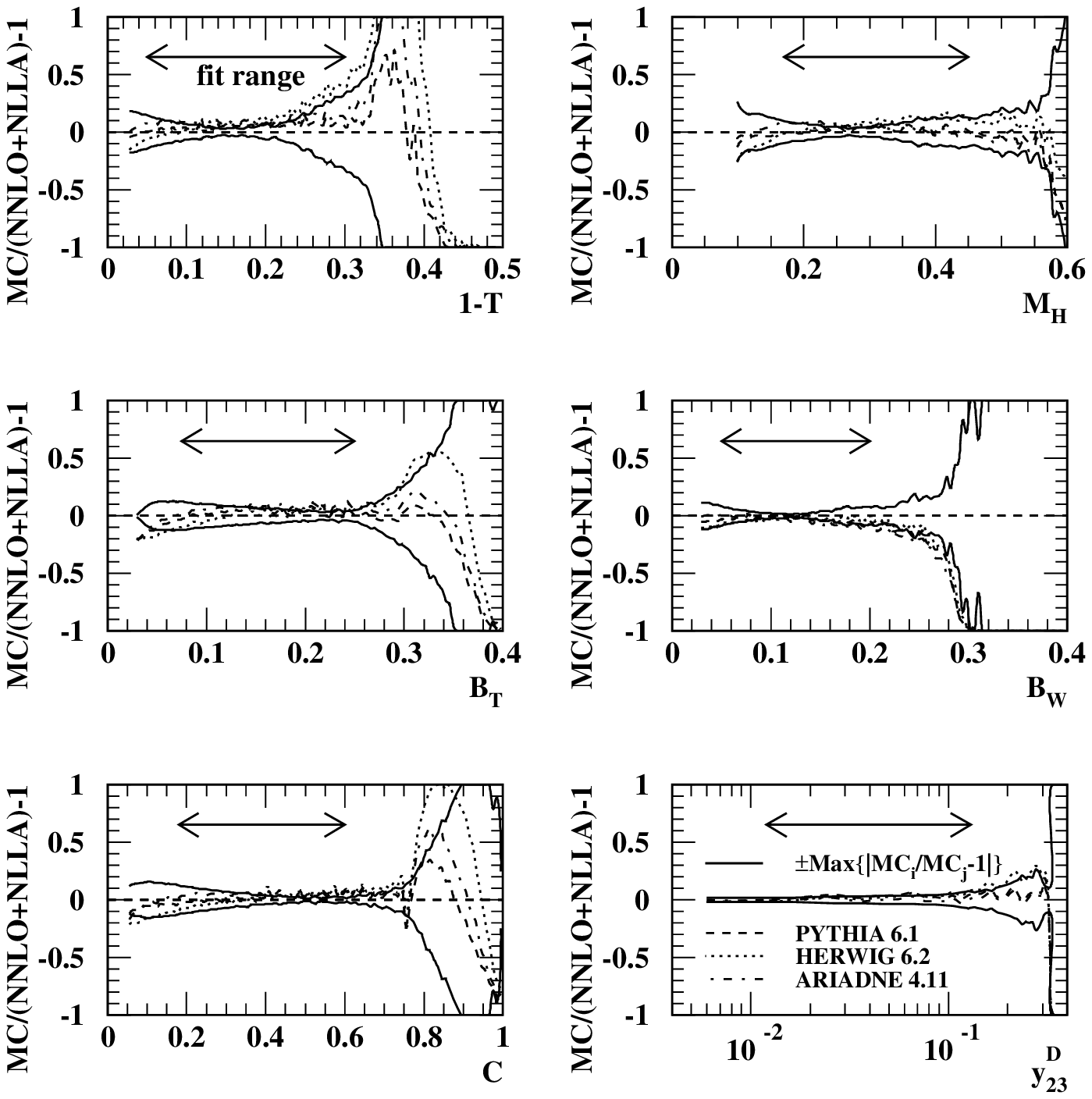}  %mctheo2_91.0_NNLO+NLLA.eps 
\caption{Comparison of NNLO+NLLA calculations for the \thr, \mh, \bt, \bw, \cp\ and \ytwothree\
  variables with the parton level predictions of the Monte Carlo generators \Pythia~6.1, \Herwig~6.2 and \Ariadne~4.11
  at \rs=91~GeV  
  (see Sect.~\ref{MeasurementsAs}).
  The vertically symmetrical band between solid lines shows the maximum 
    deviation of the ratio from one, between the three 
    generators MC$_i$ in the positive and the negative direction.
 The arrows indicate the respective fit ranges.}
\label{McVsCalc}
\end{center}
\end{figure*}

We compare the parton-level
calculations of the Monte Carlo generators with the QCD calculations in 
NNLO+NLLA with $\asmz=0.118$, $\xmu=1.0$ and $\xL=1.0$ at $\roots=91$~GeV,
see Fig.~\ref{McVsCalc} and \cite{jadeNNLO}.
  The dashed line shows the deviation from one of the ratio 
  between the Pythia prediction at the parton level and the 
  NNLO+NLLA calculation.  The dotted (dash-dotted) line shows
  the corresponding result from Herwig (Ariadne).
  The solid lines, symmetric about zero, show the maximum
  difference between any pair of the Monte Carlo models, which
  are seen to be of similar size to the differences between
  the generators and the NNLO+NNLA calculation. Therefore we consider that the
   MC simulations adequately account for the hadronization
   correction to the calculations.

The model
dependence of the hadronization correction 
is included as a systematic uncertainty as described below.

\section{Experimental Procedure\label{sec_anal}}
The data used in the present paper are identical to those
presented in~\cite{OPALPR404,OPALjets,VierjetOPAL}. For
completeness 
and to facilitate the discussion of systematic
uncertainties, we give a brief 
summary of the analysis procedure below.

\subsection{Event Selection} 
The event selection procedure is described in~\cite{OPALPR404}.  This procedure selects
well-measured hadronic event candidates, removes events with a large amount of initial-state radiation 
(ISR) at 130~GeV and above, 
and removes four-fermion background events
above the \ww\ production threshold of 160~GeV.

\subsection{Corrections to the Data}\label{detcorr} 
For each accepted event, the value of each event shape observable
is computed.  
A standard algorithm~\cite{OPAL-Higgspaper} is applied
to mitigate the
double-counting of energy between the tracking chambers and calorimeter.
To correct for background contributions,
the expected number of residual four-fermion  
events $b_i$
is subtracted from the number of data events $N_i$
in each bin $i$ of each distribution.  
Simple bin-by-bin corrections are applied to account for detector acceptance,
resolution, and the effects of residual ISR.
We examine the Monte Carlo predictions for the event shape variables at two levels:
the detector level, which includes simulation of the detector and the same analysis
procedures as applied to the data, and the hadron level, which uses stable particles\footnote{ 
For this purpose, all particles having proper lifetimes
greater than $3\times10^{-10}$~s are regarded as stable.}
only, assumes perfect reconstruction of the particle momenta, and requires the hadronic
c.m.\ energy $\sqrt{s'}$ to satisfy $\sqrt{s}-\sqrt{s'}<1$~GeV. 
The ratio of the hadron- to the detector-level prediction for each bin, 
$\alpha_i$, is used as the correction factor for the data, yielding the corrected
bin content $\widetilde{N}_i=\alpha_i(N_i-b_i)$.  
This corrected hadron-level distribution is then normalized 
by the total number of events $N=\sum_k\widetilde{N}_k$ and bin width $W_i$:
$P_i=\widetilde{N}_i/(N W_i)$. 

\subsection{Systematic uncertainties }\label{sec_syserr}
To evaluate systematic uncertainties, 
the analysis is repeated after modifying the selection or correction procedures.
The 
event and track selection cuts are varied within the ranges of values suggested in~\cite{OPALPR404}.
The Monte Carlo model employed to calculate the detector correction is altered,
and the cross section used in the subtraction of four-fermion events is varied by $\pm5\%$. 
In each case, the difference in each bin 
with respect 
to the standard analysis is taken as a contribution to the systematic uncertainty.

\section{Determinations of \boldmath{\as}\label{sec_results}}
\subsection{Fit procedure}\label{MeasurementsAs}
The strong coupling \as\ is determined using a minimum-$\chi^2$ fit, comparing theory with each of the
measured event shape distributions at the hadron level.  
A \chisq\ value is calculated at
each c.m.\ energy:
\begin{equation}
  \chisq= \sum_{i,j}^{n} (P_i-t_i(\as)) (V^{-1})_{ij} (P_j-t_j(\as))
\end{equation}
where $i,j$ 
include the bins within the fit range of the event shape
distribution, $P_i$ is the measured value in the $i$th bin, 
$t_i(\as)$ is the QCD prediction for the $i$th bin corrected for
hadronization effects, 
and $V^{-1}$ is the inverse of the statistical covariance matrix $V^{\rm stat}$ of the values $P_i$.
The QCD prediction is obtained by integrating the result in~(\ref{NNLOcalc}) over the bin
width, and then applying the hadronization correction. 
The \chisq\ value is minimized with
respect to \as\ with the renormalization scale factor $\xmu$ and the rescaling variable $x_{_L}$ set to
1.  The evolution of the strong coupling \as\ as
a function of the renormalization scale is implemented
to three-loop order~\cite{ESW}.
Since 
the c.m.\ energy range considered here does not 
cross flavour thresholds,  
no significant uncertainties are introduced by the evolution of \as.
Separate fits are performed to each of the six observables
at each c.m.\ energy value.   

To account for correlations between bins
in the computation of $\chi^2$,
the statistical covariance matrix $V_{ij}^{\rm stat}$ is 
calculated following the approach
described in~\cite{OPALPR404}:
\begin{eqnarray}
V_{ij}^{\rm stat} & = & \sum_k \frac{\partial P_i}{\partial N_k}    \label{statCov}
                \frac{\partial P_j}{\partial N_k} N_k \\ \nonumber
       & = & \frac{1}{N^4} \sum_k \alpha_k^2 N_k
         \left(N\delta_{ik} - \widetilde{N}_i\right)
         \left(N\delta_{jk} - \widetilde{N}_j\right) \;.
\end{eqnarray}
where $\delta_{ik}$ is the Kronecker delta function, and the other terms are defined in Sect.~\ref{detcorr}.

To test our procedure, we verified that we are able to reproduce
the fit results from~\cite{OPALPR404,MAF} using NLO+NLLA predictions within typically 0.5\%.

The fit ranges are chosen as in~\cite{OPALPR303,OPALPR404} and correspond to regions where the 
corrections of the data 
and of the parton level theory to the hadron level
are both
reasonably small and where the NLO+NLLA fit results are stable under
small variations of the fit range. The relevant
calculations in the two-jet region are the NLLA terms, which are identical in the present study 
and~\cite{OPALPR404}.
The NNLO prediction for \ytwothree\ is given in bins of $-\ln \ytwothree$.  
The transformation to bins of \ytwothree\ produces 
precision problems for small values of \ytwothree. Therefore we use a more
restrictive fit range for \ytwothree\ than in~\cite{OPALPR404},
which approximately matches the range in~\cite{jadeNNLO}: the lower bound is
0.012 rather than 0.0023.

The fit ranges are shown in Tab.~\ref{fitranges}.
In the resummation of log-enhanced terms, 
the leading log term of ${\rm d}A/dy$ is $\ln(y)/y$. 
At $y$ values equal to the lower limit $y_0$, the value of
 $\ash\ln(y_0)/y_0$ is still of the order of one 
for $\asmz=0.118$ and $91\le\roots\le207$~GeV.
Thus the fit ranges are also suitable for pure NNLO analysis. 

Previously measured event shape distributions are published~\cite{OPALPR404} only in the
energy ranges as separated by horizontal lines in Tab.~\ref{TableNumberEvents}.
We perform fits to extract \as\ at all of the energy 
points\footnote{This allows 
the use of
the event shape covariance matrices which are energy dependent and have been 
calculated separately for each energy point.} 
shown in Tab.~\ref{TableNumberEvents} column 2.  We average over the larger energy ranges 
only for purposes of presentation.

The detector correction factors are typically
between 0.9 and 1.5 within the fit ranges. 
For energies above 189~GeV,
which have low statistics, 
the corrections exceed 2.0 in the multi-jet regions.
The maximum correction of 4.8 occurs for the \thr\ distribution at 207~GeV.
The variable with the least variation in the size of its hadronization
correction, and with hadronization corrections closest to one, is \ytwothree.

The statistical uncertainty of 
  \as\ is given by the variation 
  required to increase \chisq\ by one unit from its minimum. The systematic
  uncertainties account for
experimental effects, the ha\-dronization 
correction procedure and uncertainties of the theory.  The
three sources of systematic uncertainty are added in quadrature
to the statistical uncertainty 
  to define the total uncertainty. 
Below, we describe the evaluation of systematic uncertainties.
For each variant of the 
analysis, the corresponding distribution is fitted 
to determine \as, and the difference with respect to 
the value of \as\ from 
the default analysis is taken as a systematic uncertainty contribution.
\begin{description}
\item[Experimental Uncertainties:] 
  These are assessed as
described in Sect.~\ref{sec_syserr}.   
\item[Hadronization:] For the default analysis, PYTHIA is used to
    evaluate hadronization corrections 
  (Sect.~\ref{qcdprediction}).  As systematic variations, HERWIG and
  ARIADNE are used  
  instead. The larger
  of the deviations is taken as the systematic 
  uncertainty.
  It was observed in~\cite{jones03,pedrophd} that systematic
  uncertainties 
    determined from the differences between the PYTHIA, HERWIG, and ARIADNE
  models are generally much larger than
    those that arise 
  from varying the
  parameters of a given model.
\item[Theoretical Uncertainties:] The theoretical calculation of 
event
  shape observables is a finite power series in \as.  The
  uncertainties originating from missing higher order terms are
  assessed by changing the renormalization scale factor to $\xmu=0.5$
  and $\xmu=2.0$. The rescaling variable is set to $\xL=2/3$ and $\xL=3/2$ ($\xL=4/9$ and $\xL=9/4$ in
  case of \ytwothree)~\cite{OPALPR404}. 
  The largest deviation 
    with respect to the standard analysis 
  is taken as the systematic uncertainty.
  A variation of the matching scheme is not studied because $R$-matching is not available for NNLO+NLLA.
  The matching scheme was not an important source of uncertainty
  in studies of $\as$ based on NLO+NLLA calculations.

\end{description}

\subsection{Results from NNLO Fits}
The results of the NNLO fits are summarised in
Tab.~\ref{asresultsnnlo}.  
In order to clarify the presentation, 
the \as\ values from 130 and 136~GeV are 
combined\footnote{This combination is performed 
assuming minimum overlap correlation 
of the experimental uncertainties at the different c.m.\ energy points.
The procedure includes a correction for the running of $\as(\roots)$ to the
luminosity weighted mean energy.
}, 
and likewise the values from 161 to 183~GeV, and from 189~GeV and above. These 
three ranges of c.m.\ energies, which correspond to luminosity weighted mean energies
of 133.1, 177.4 and 197.0~GeV, respectively, cover sufficiently small ranges of \rs\ 
that only small variations of \as\ occur within each range.

\begin{figure}[htb!]
\includegraphics[width=.5\columnwidth]{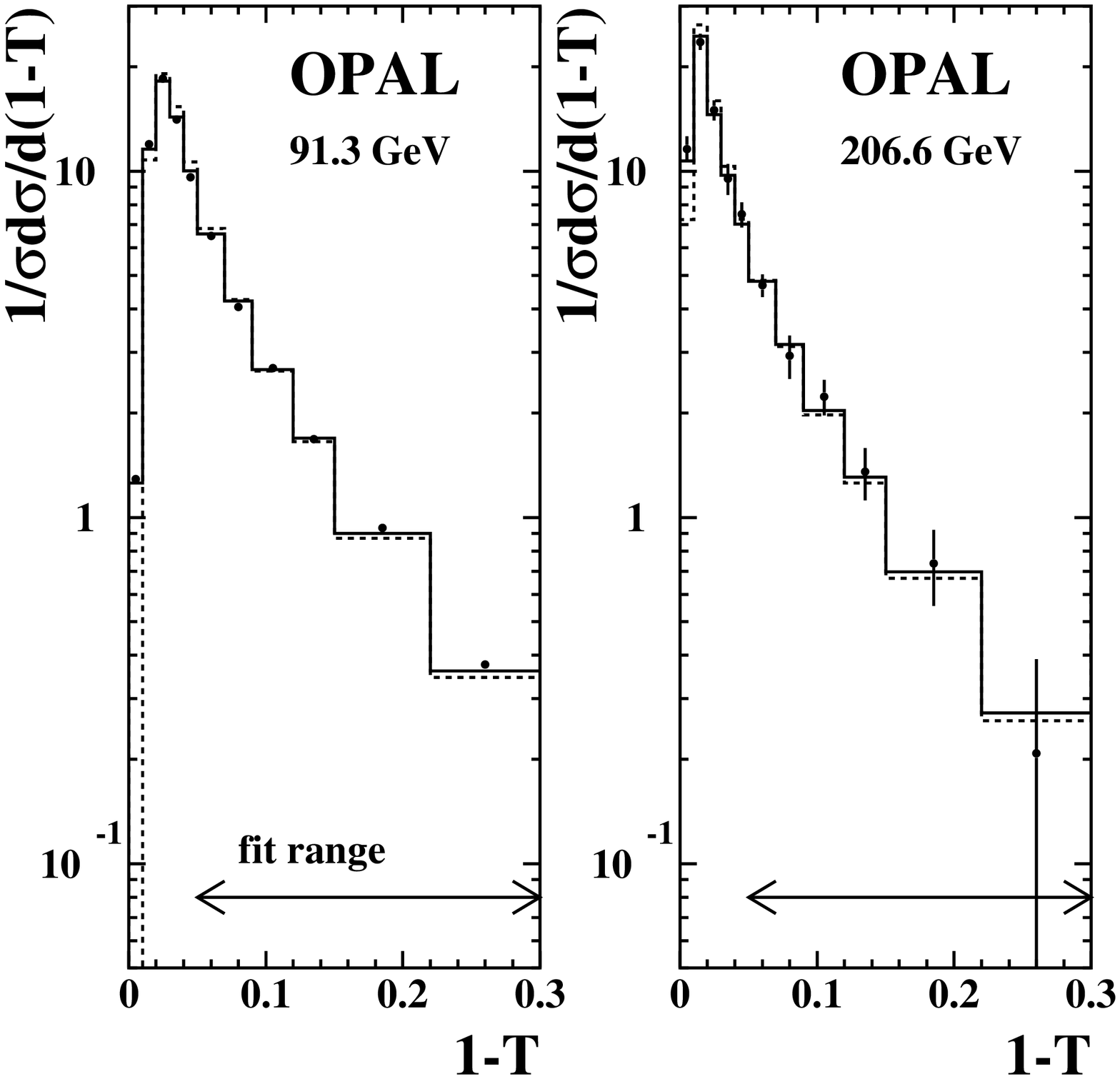}  % eventshape_fit_1_1,13_17,18.eps
\includegraphics[width=.5\columnwidth]{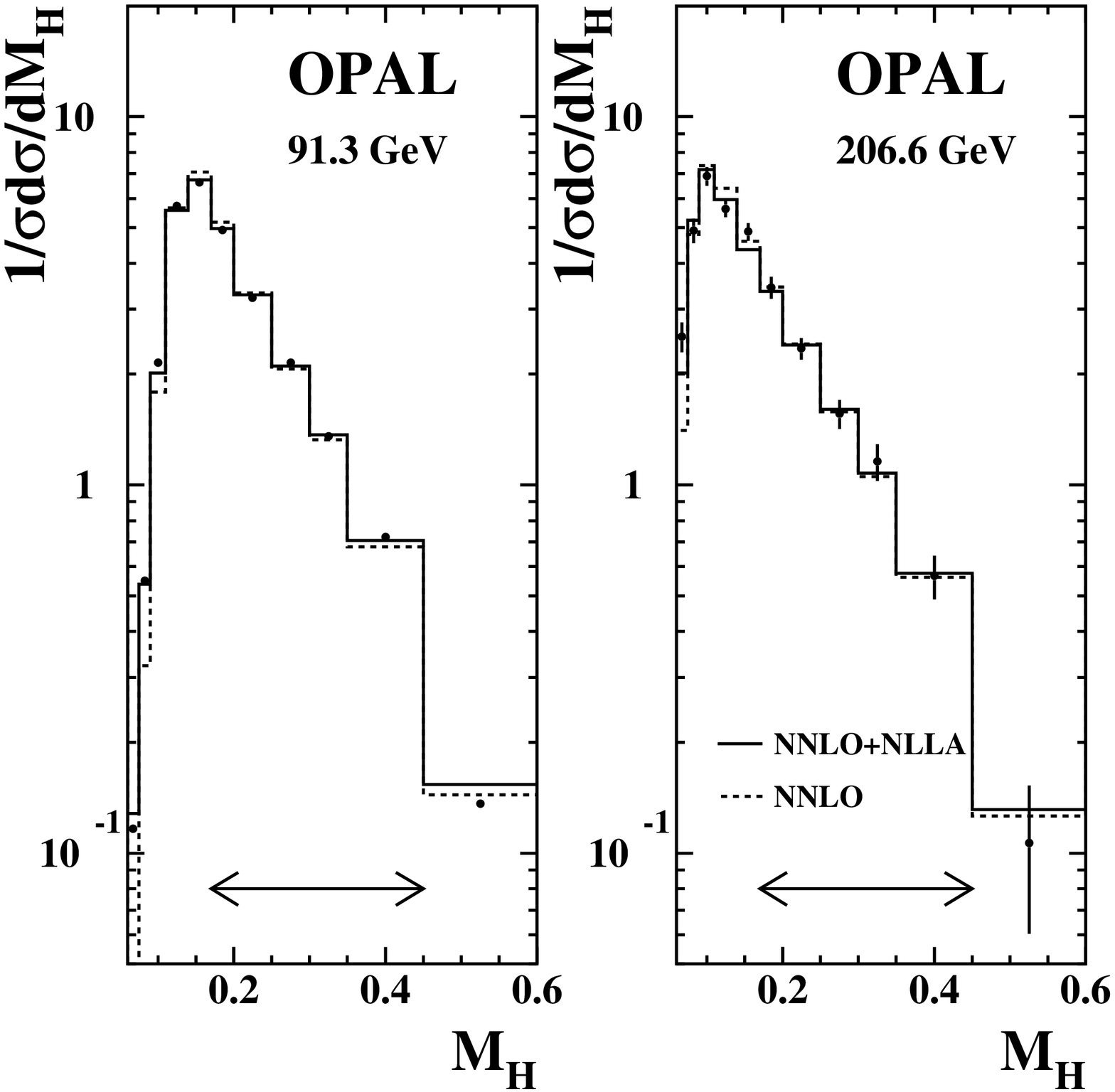}  % eventshape_fit_2_1,13_17,18.eps
\caption{The points show the \thr\ and \mh\ 
distributions at the hadron level for
$\roots=91.3$ and 206.6~GeV with statistical uncertainty bars. Some
uncertainty bars are smaller than the data points.  Superimposed as
histograms are the NNLO and NNLO+NLLA predictions combined with hadronisation
effects using the corresponding fit results for $\as(\roots)$ shown in
Tabs.~\ref{asresultsnnlo},~\ref{asresultsnnlonlla}.  The arrows indicate the fit ranges.}
\label{esdistributions1}
\end{figure}

\begin{figure}[htb!] 
\includegraphics[width=.5\columnwidth]{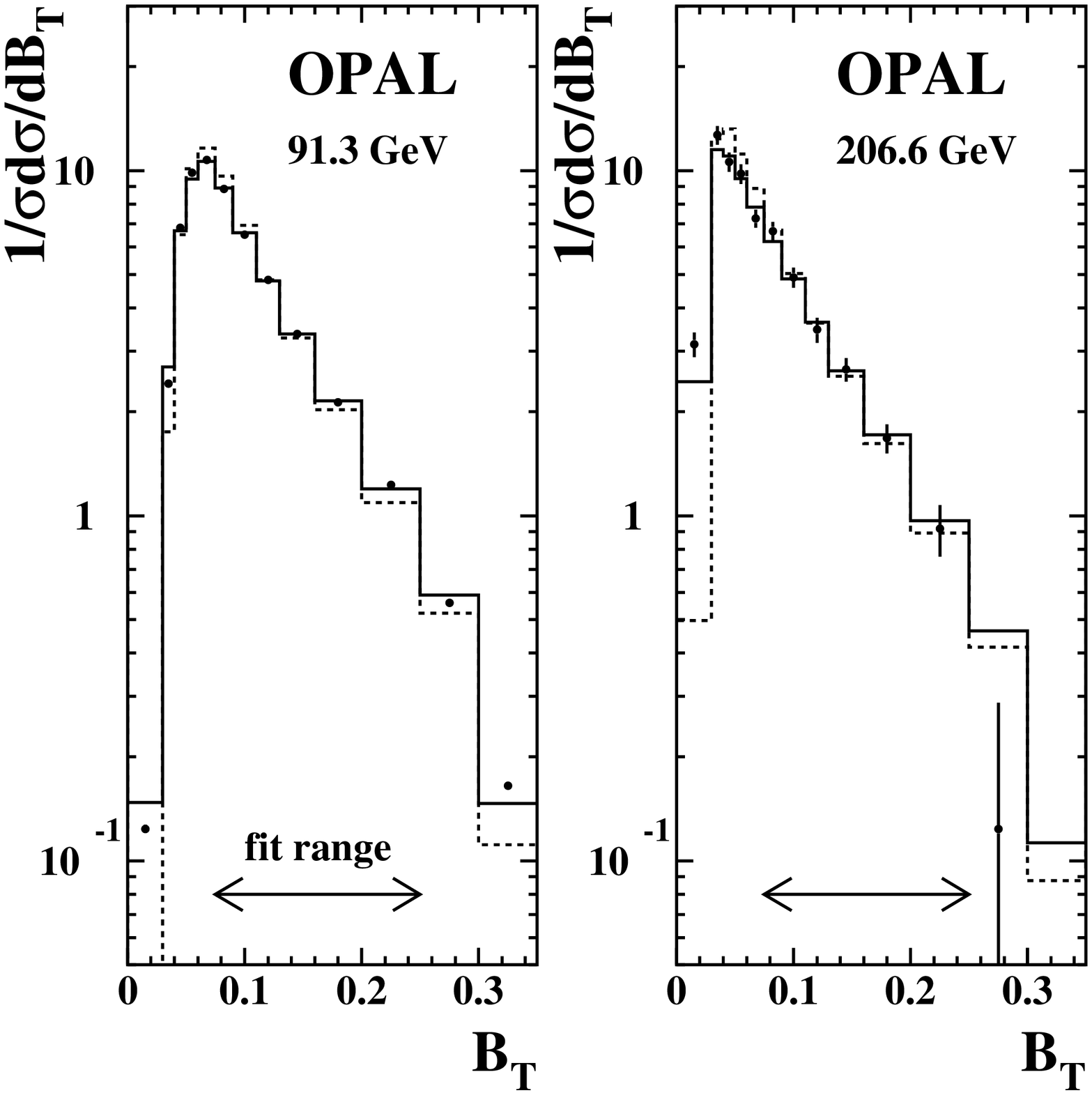}  % eventshape_fit_3_1,13_17,18.eps 
\includegraphics[width=.5\columnwidth]{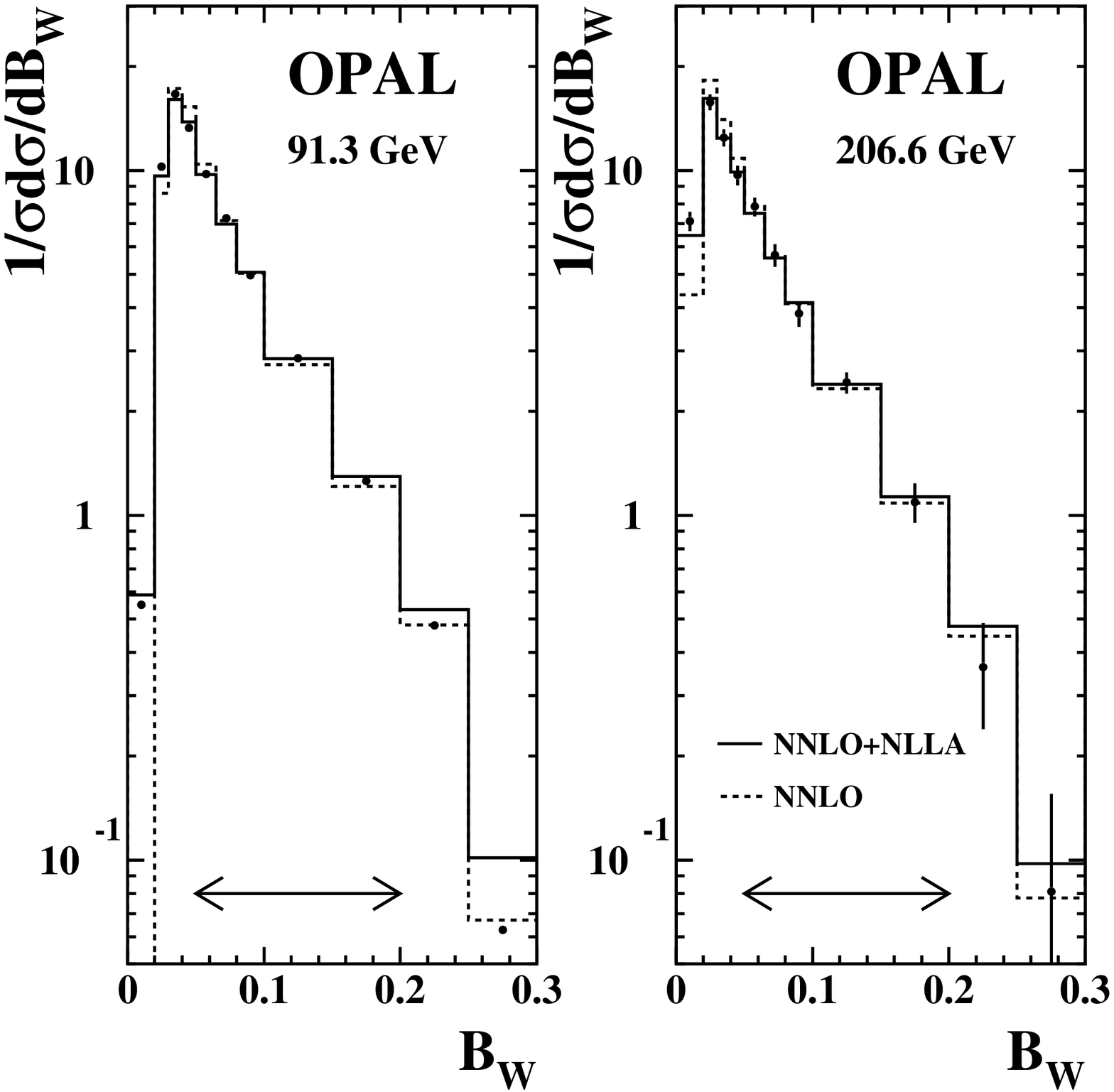}  % eventshape_fit_4_1,13_17,18.eps
\caption{The points show the \bt\ and \bw\ distributions at the hadron level for 
$\roots=91.3$ and 206.6~GeV with statistical uncertainty bars. Some
uncertainty bars are smaller than the data points.  Superimposed as
histograms are the NNLO and NNLO+NLLA predictions combined with hadronisation
effects using the corresponding fit results for $\as(\roots)$ shown in
Tabs.~\ref{asresultsnnlo},~\ref{asresultsnnlonlla}.  The arrows indicate the fit ranges.}
\label{esdistributions2}
\end{figure}

\begin{figure}[htb!]
\includegraphics[width=.5\columnwidth]{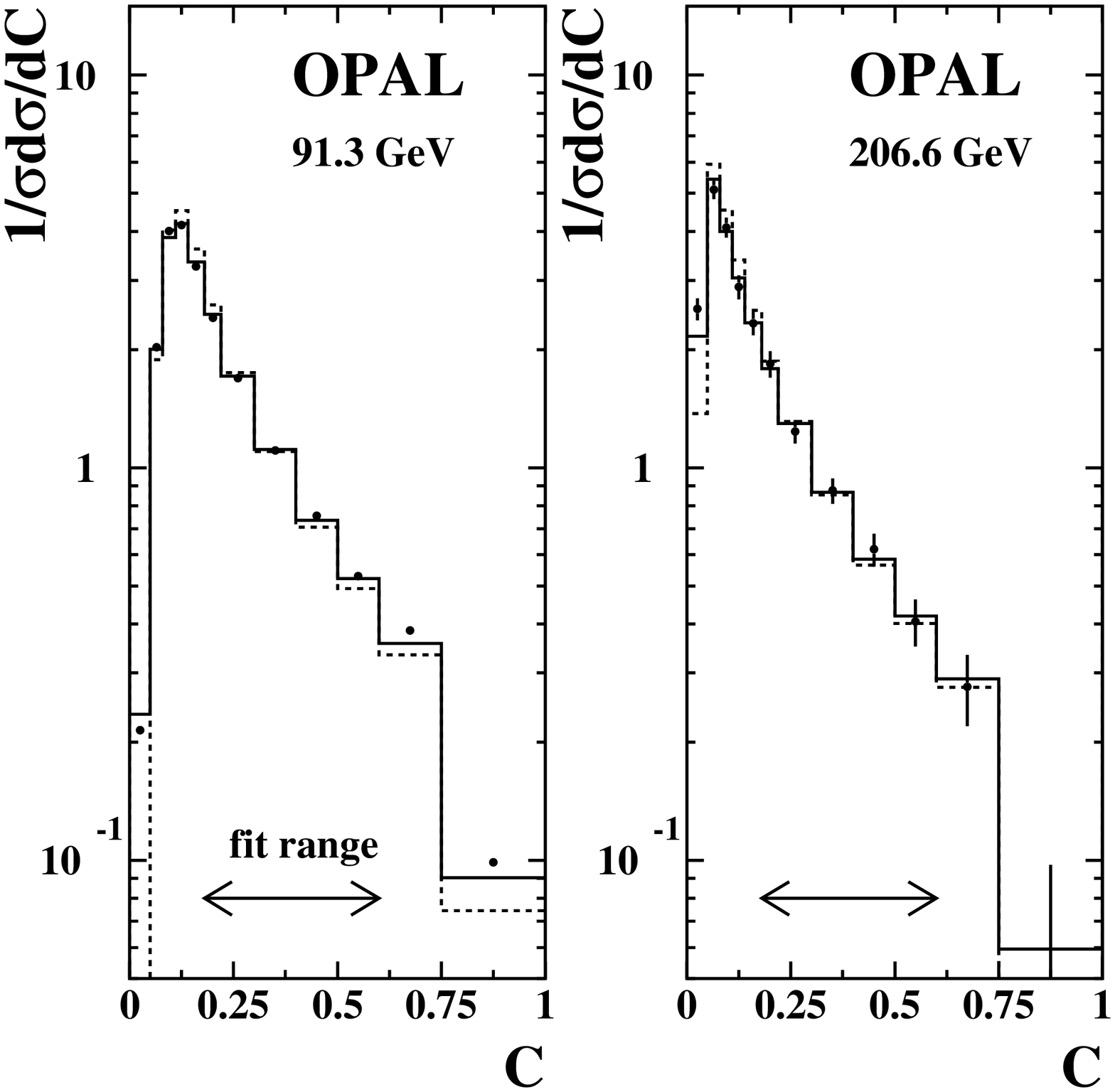}  % eventshape_fit_5_1,13_17,18.eps
\includegraphics[width=.5\columnwidth]{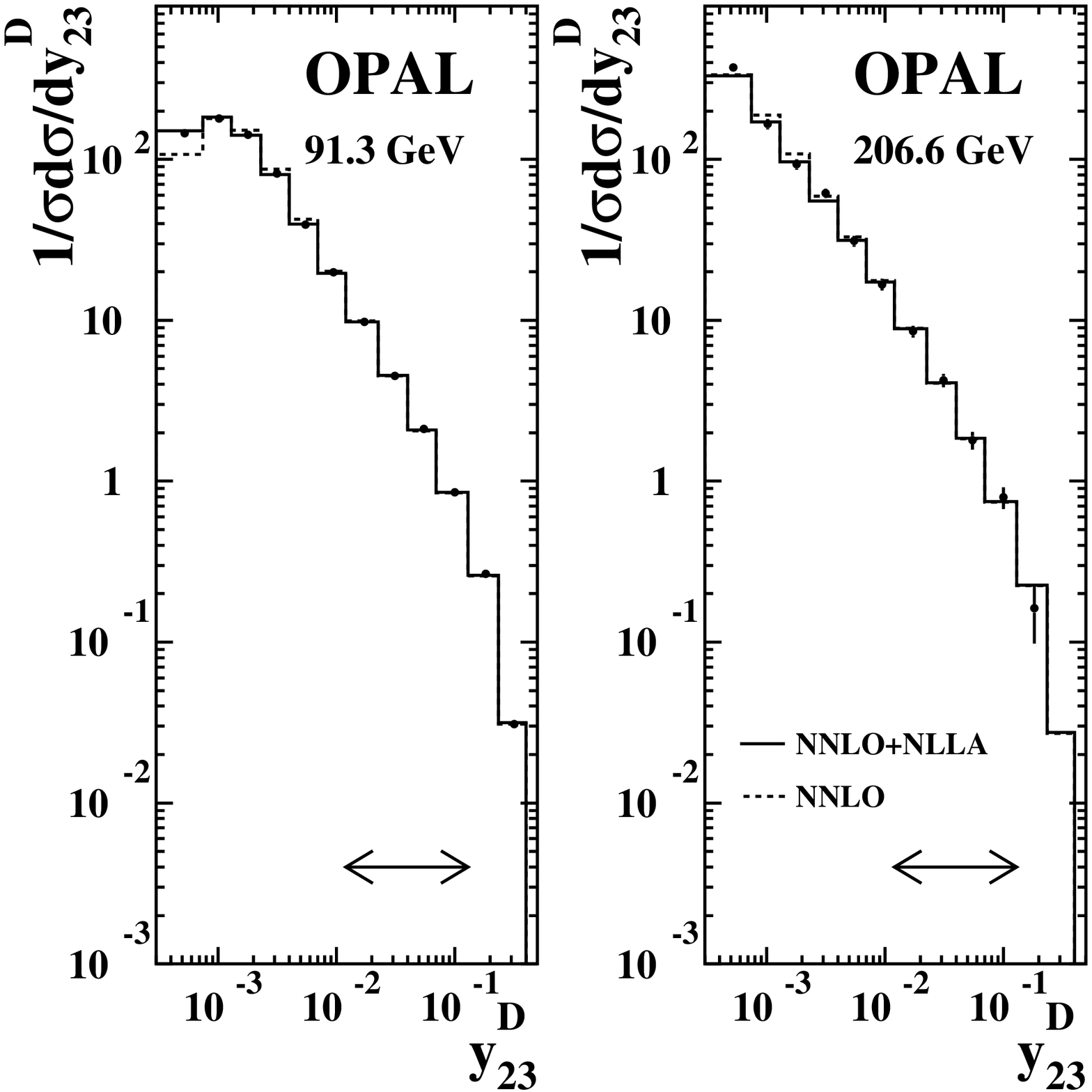}  % eventshape_fit_6_1,13_17,18.eps
\caption{The points show the \cp\ and \ytwothree\ distributions at the hadron level for 
$\roots=91.3$ and 206.6~GeV with statistical uncertainty bars. Some
uncertainty bars are smaller than the data points.  Superimposed as
histograms are the NNLO and NNLO+NLLA predictions combined with hadronisation
effects using the corresponding fit results for $\as(\roots)$ shown in
Tabs.~\ref{asresultsnnlo},~\ref{asresultsnnlonlla}.  The arrows indicate the fit ranges.}
\label{esdistributions3}
\end{figure}

Figs.~\ref{esdistributions1}, \ref{esdistributions2} and \ref{esdistributions3} show the 
\thr, \mh, \bt, \bw, \cp\ and \ytwothree\
event shape distributions together with the NNLO fit results for the 
91.3 and 206.6~GeV energy points. 
All event shape distributions are described reasonably 
  well
by the fitted predictions.
The \chisqd\ values, which are large for the data at the \znull\ peak 
because they are based on statistical uncertainties only, range from
0.03 for \ytwothree\ at
$\roots=191.6$~GeV to 228 for \bt\ at $\roots=91.3$~GeV. 
The uncertainties at this point are dominated by the experimental systematic uncertainties (discussed below). 
To check how the inclusion of the experimental errors in the fits
reduces the large \chisq\ values, an alternative \chisq\ value
is defined as follows. An
estimate of the experimental systematic covariance is added to the statistical covariance~(\ref{statCov}),
\begin{equation}
  V_{ij}^{\rm total}  =  V_{ij}^{\rm stat} + \min(\sigma_{{\rm exp},i}^2, \sigma_{{\rm exp},j}^2), \label{totCov}
\end{equation}
where $\sigma_{{\rm exp},i}$ is the experimental systematic uncertainty at bin $i$.
Tab.~\ref{NNLOminOvl} shows the fit results at 91~GeV.
The fit results are in general not
compatible within the combined statistical and experimental errors with the results when
only the statistical covariance is used.
 These \chisqd\ values are smaller
than those from employing the statistical covariances,
however they still range up to
62 for \bt. 
We return to this issue when we discuss the NNLO+NLLA fits below.

  The results for \as\ from the different event shape variables
  are remarkably consistent with each other, for each c.m.\ energy,
  as seen from 
Tabs.~\ref{detailedasresultsnnlo},~\ref{detailedasresultsnnlo2}
(but not in the case of the modified 
fit procedure whose results are shown in Tab.~\ref{NNLOminOvl}).
We find root-mean-square (r.m.s.) values for
$\as(\rs)$ between 0.0012 at 183~GeV and 0.0044 at 202~GeV.  
The variations in \as\ between the different variables are comparable to the total systematic uncertainties.
The values of \as\ are significantly larger at 91 GeV than at
higher energies, providing evidence for the running of \as.  
  Theoretical uncertainties dominate at the \znull~peak (except for \ytwothree),
  where the statistical uncertainties are small. Similarly, statistical
  uncertainties are small at 189 GeV where there is a relatively 
  large data sample.  Statistical uncertainties dominate at 130, 161 
  and 172~GeV, where the data samples are smaller.  Experimental 
  systematic uncertainties are small
at 91~GeV and larger at higher \rs\ where uncertainties from
subtraction of ISR and four fermion events contribute more strongly.
For completeness, the fit results for all energy points are given separately
in App.~\ref{separateFits}.

\subsection{Results from NNLO+NLLA Fits}
The results of the NNLO+NLLA fits are shown in Figs.~\ref{esdistributions1}, \ref{esdistributions2} and \ref{esdistributions3} and 
listed in Tab.~\ref{asresultsnnlonlla}.  
The calculations fit the data better than the pure NNLO calculations at small
values of the variables. 
  The values of \chisqd\ are smaller on average than for the NNLO fits,
  especially at 91~GeV, indicating better 
  consistency with the data.
The r.m.s.\ values of $\as(\roots)$ 
  vary
between 0.0017 at $\roots=205$~GeV and 0.0051 at $\roots=202$~GeV, i.e.\
the scatter of individual results is essentially the same as for the
NNLO analysis.  The pattern of statistical, experimental, and
hadronization uncertainties is the same as for the NNLO fits discussed
above. 
At most energy points the theory error results from the \xL\ variation for \mh, 
\bw, \ytwothree, and from the \xmu\ variation for \thr, \bt, \cp.  
Compared with the NNLO analysis 
the values of \as\ are lower by 0.4\% on average,
and the theoretical uncertainties are 
higher by 33\%.  
Larger scale uncertainties are expected to arise at 91~GeV when the NLLA
terms are added to the NNLO terms~\cite{NNLONLLAES,NNLONLLAEStalk,alephNNLONLLA}, because the
NNLO calculation compensates for the variation of the renormalization scale in two loops, while the NLLA term
compensates for the variation in only one loop. 
  The difference we observe between \as\ in the NNLO and NNLO+NLLA
  studies is smaller than the corresponding difference observed between
  the NLO and NLO+NLLA studies (Sect.~\ref{compNLONLLA}), as predicted in~\cite{NNLONLLAES}.  
  This difference is also smaller than that observed at lower 
  energies~\cite{jadeNNLO}, as expected from the energy evolution of \as.

The fit results at 91~GeV employing the total covariance~(\ref{totCov}) are shown in
Tab.~\ref{NNLONLLAminOvl}. Unlike the case of pure NNLO, the fit results are compatible
with the results based on only the statistical covariance within the combined statistical and 
experimental systematic uncertainties. The \chisqd\ values are of the order of 1. 
For a reasonable description of the data by the theory
predictions, inclusion of the resummed logarithmic terms (NLLA) is seen to be
important. As the covariance~(\ref{totCov}) is only an approximation, we do not use the
fit values or errors further.

\subsection{Combination of Results\label{combination}}
\begin{figure}[htb!]
\begin{center}
\includegraphics[width=1.\columnwidth]{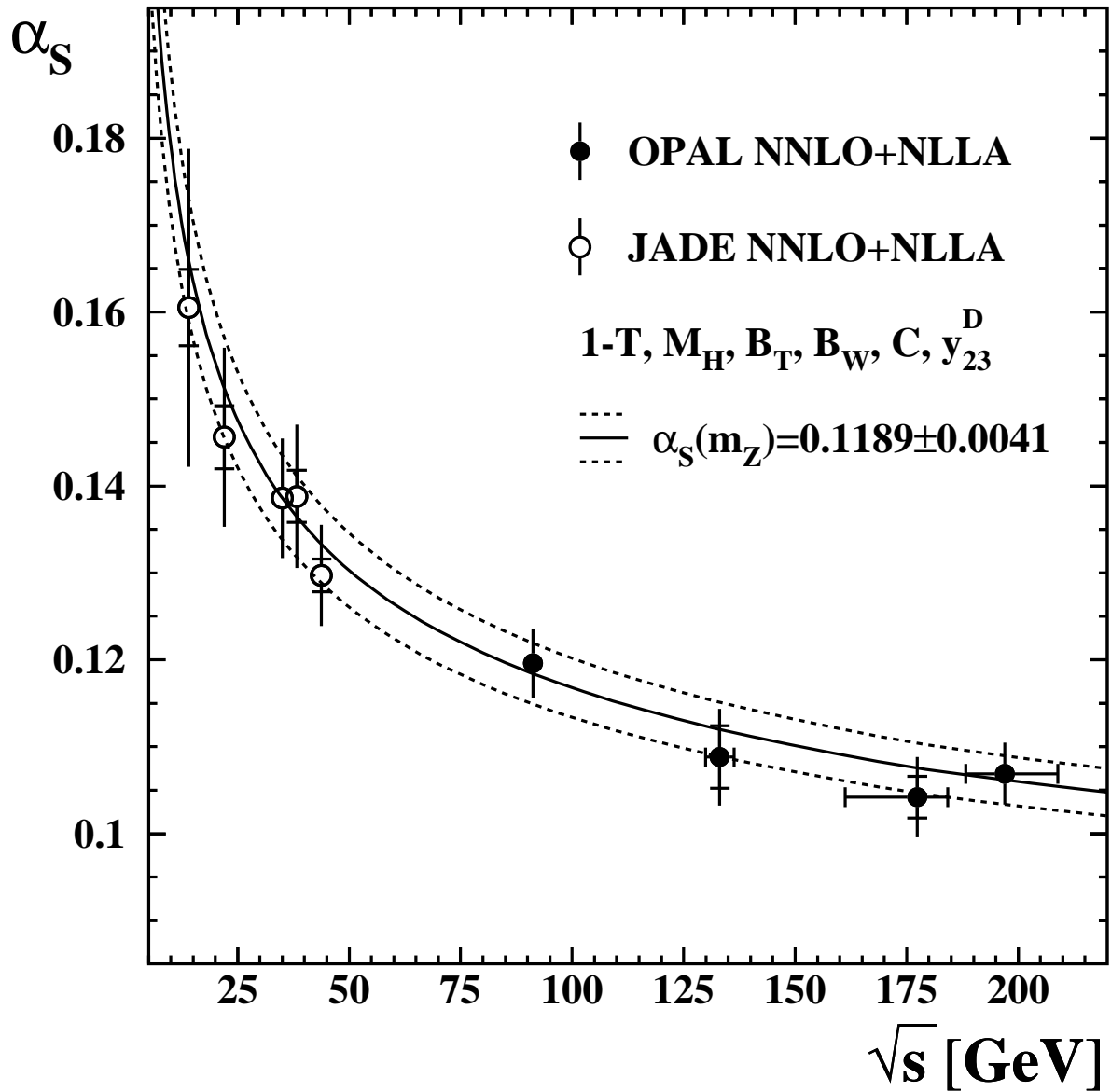}  % asplotopal.eps
\caption{The points show the values of \as\ for the OPAL energy ranges.  
The inner
uncertainty bars show the combined statistical and experimental
uncertainties and the outer the total uncertainties.  The
full and dashed lines indicate the \as\ result from the NNLO+NLLA analysis
that combines all variables and OPAL energy points.
The results from the NNLO+NLLA analysis of JADE
data~\cite{jadeNNLO} are shown as well.}
\label{asvscme}
\end{center}
\end{figure}

The results obtained at each energy point for the six event shape
observables are combined using uncertainty-weighted averaging as
in~\cite{MinovlConf,OPALPR404,ALEPH,STKrev}.  The statistical
correlations between the six event shape observables are estimated at
each energy point from fits to hadron-level distributions derived from
50 statistically-independent Monte Carlo samples.  The experimental
uncertainties are determined assuming that the smaller of a pair of
correlated experimental uncertainties gives the size of the fully correlated
uncertainty. The minimum overlap assumption results in a conservative estimate
of the total uncertainty.  The hadronization and theoretical systematic
uncertainties are evaluated by repeating the combination with changed
input values, i.e.\ using a different hadronization model or the
different value of \xmu\ or \xL\ that yields the maximum deviation.  The results are given in
Tab.~\ref{asecmcomb} and shown for the NNLO+NLLA analysis in
Fig.~\ref{asvscme}.
Also shown is a
comparison with the results of 
a NNLO+NLLA analysis at lower energy from the JADE Collaboration~\cite{jadeNNLO}.

To study the compatibility of our data with the QCD
prediction for the evolution of the strong coupling with c.m.\ energy we
repeat the combinations with or without evolution of the combined
results to the common scale, setting the theory uncertainties to zero
since these uncertainties are highly correlated between energy points.
We assume the hadronization uncertainties to be
partially correlated.   The \chisq\ probability of the average for a running
  (constant) coupling in the NNLO+NLLA study is 0.59 
  (6$\times10^{-13}$).  The corresponding result for
  the NNLO study is 0.56 (2$\times 10^{-6}$).
We interpret this 
as clear evidence, from OPAL data alone, for the running of \as\
in the manner predicted by QCD.

We evolve the results in Tab.~\ref{asecmcomb} with $\rs>\mz$ to a common scale \mz\
using the QCD formula for the running of \as.  The results are then
combined using the uncertainty-weighted averaging procedure mentioned above.
The results are shown in Tab.~\ref{asecmcomb_all} 
for the NNLO and NNLO+NLLA analyses.
The consistency between these values and the
results from 91~GeV    
(Tab.~\ref{asecmcomb}) 
  demonstrates the compatibility of 
the data              
with the running predicted by QCD. 
The high energy data have 
smaller theoretical and hadro\-ni\-zation
uncertainties, and therefore complement the statistically superior 91~GeV data.

The values of \asmz\ obtained from combining all energy points 
  are also 
  given in Tab.~\ref{asecmcomb_all}. 
  The resulting values of 
    $\resulttotnnlo$ from the NNLO study, and
    $\resulttotnnlonlla$ from the NNLO+NLLA study,
  are consistent with the world average~\cite{bethke09} ($0.1184\pm0.0007$),
  recent NNLO analyses of JADE~\cite{jadeNNLO} ($0.1172\pm0.0051$) and ALEPH~\cite{asNNLO}
  ($0.1240\pm0.0033$) data, our previous study~\cite{OPALPR404} based on 
  NLO+NLLA calculations ($0.1191\pm0.0047$) and the study~\cite{alephNNLONLLA}
  of ALEPH data using NNLO+NLLA ($0.1224\pm0.0039$).
The total uncertainties of 2.6\% and 3.4\% for \asmz\ place these measurements amongst the 
most precise determinations of \as\ available.

\begin{figure}[htb!]
\begin{center}
\begin{tabular}[tbp]{cc}
\includegraphics[width=1.\columnwidth]{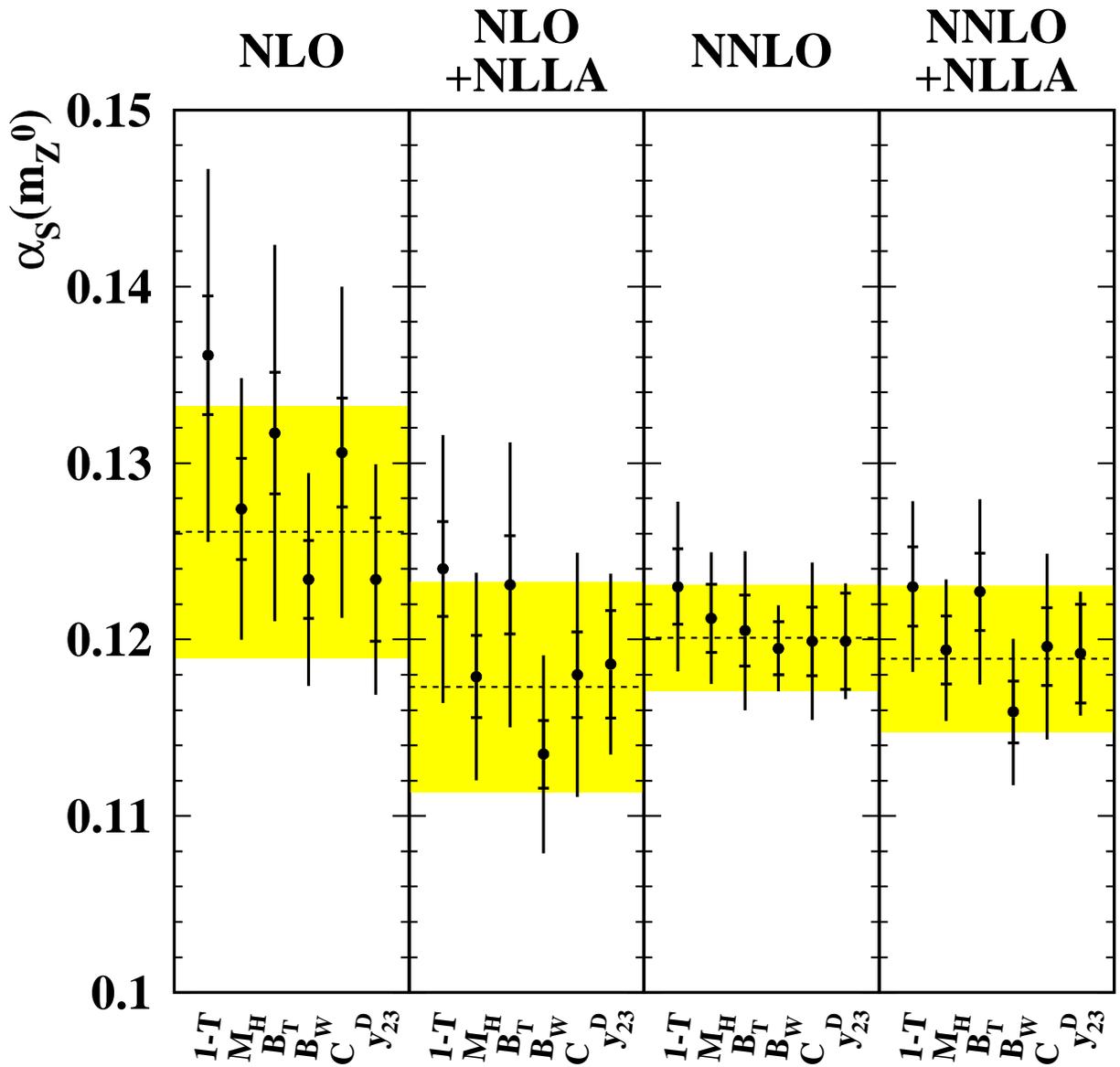}  % obsscatterplot2.eps
\end{tabular}
\caption{\as\ results combined over all OPAL c.m.\ energies for different
event shape variables and different QCD calculations as indicated
on the figure.  
The shaded bands and
dashed lines show the values of \asmz\ combined from these values with total
uncertainties.  The inner and outer uncertainty bars show the combined
statistical and experimental and total uncertainties, respectively.}
\label{asobsscatter}
\end{center}
\end{figure}

After running the fit results for $\as(\roots)$ for each observable to
the common reference scale \mz, we combine the results for a given
observable to a single value.  We use the same method as above and
obtain the results for \asmz\ shown in Tab.~\ref{asvarcomb}. 
  These results demonstrate that
the measurements from the different observables are far from compatible 
  with each other
when only statistical uncertainties are
considered, but are consistent with a common mean when 
the systematic uncertainties are included, neglecting their correlation.
The r.m.s. values 
of the results for \asmz\ are 
0.0013 for the NNLO analysis and
0.0026 for the NNLO+NLLA analysis; both values lie within the range of the uncertainty for the corresponding combined result shown in
Tab.~\ref{asecmcomb_all}.
Fig.~\ref{asobsscatter} displays
the combined \as\ result for each observable.
Results from NLO and NLO+NLLA studies, discussed below, are also shown.
Combining the NNLO or NNLO+NLLA results of Fig.~\ref{asobsscatter} to obtain an overall
value for \as, or evolving each event shape measurement from each
energy to the reference scale and then combining, yields the same result
as given by the corresponding measurement in Tab.~\ref{asecmcomb_all} to within
  $\Delta\asmz=0.0003$.

\subsection{Comparison with NLO and NLO+NLLA fits}
\label{compNLONLLA}
To compare our results with previous \as\ measurements, the
fits to the event shape distributions are repeated with NLO
predictions and with NLO predictions combined with resummed NLLA with
the modified $\ln R$-matching scheme (NLO+NLLA), both with $\xmu=1$.
The NLO+NLLA calculations with the modified $\ln R$-matching scheme
were the standard of the 
analyses at the time of termination of the LEP experiments~\cite{L3,ALEPH,OPALPR404,delphi327}.  
  The fit ranges and procedures for the evaluation of systematic
  uncertainties are the same as those used above for the NNLO and
  NNLO+NLLA studies and thus differ somewhat from the previously
  published results~\cite{OPALPR075,OPALPR404}.

The combination of the fits using NLO predictions %%returns -- Gary: 
yields
$\asmz=0.1261$\linebreak$\pm0.0011\stat\pm0.0024\expt\pm0.0007\had\pm0.0066\theo$
while the combination of\linebreak NLO+NLLA results yields
$\asmz= 0.1173$ 
$\pm0.0009\stat\pm0.0020\expt\pm0.0008\had$\linebreak$\pm0.0055\theo$.
These results are shown 
  by the corresponding shaded bands 
in Fig.~\ref{asobsscatter}.
The result obtained with the NLO+NLLA prediction is consistent with
the NNLO and NNLO+NLLA analyses, but the theory uncertainties are
larger by a factor of 2.3 and 1.5 respectively.  The analysis using NLO predictions gives
theoretical uncertainties larger by a factor of 2.8 and 1.8, and the value for
\asmz\ is larger compared to the NNLO or NNLO+NLLA results.  It has
been observed previously that values for \as\ from NLO analyses with
$\xmu=1$ are large in comparison with most other
analyses~\cite{OPALPR054}.  
The NLO+NLLA analysis  
yields a smaller value of \asmz\ compared to the NLO result, and the NNLO+NLLA analysis a %larger 
smaller one compared to NNLO.
The difference between the NNLO+NLLA and NNLO results is smaller than
the difference between those of the NLO+NLLA and NLO fits because a larger part of the
NLLA terms is included in the NNLO calculations.

\subsection{Renormalization Scale Dependence}
\begin{figure}[htb!]
\begin{center}
\includegraphics[width=0.65\columnwidth]{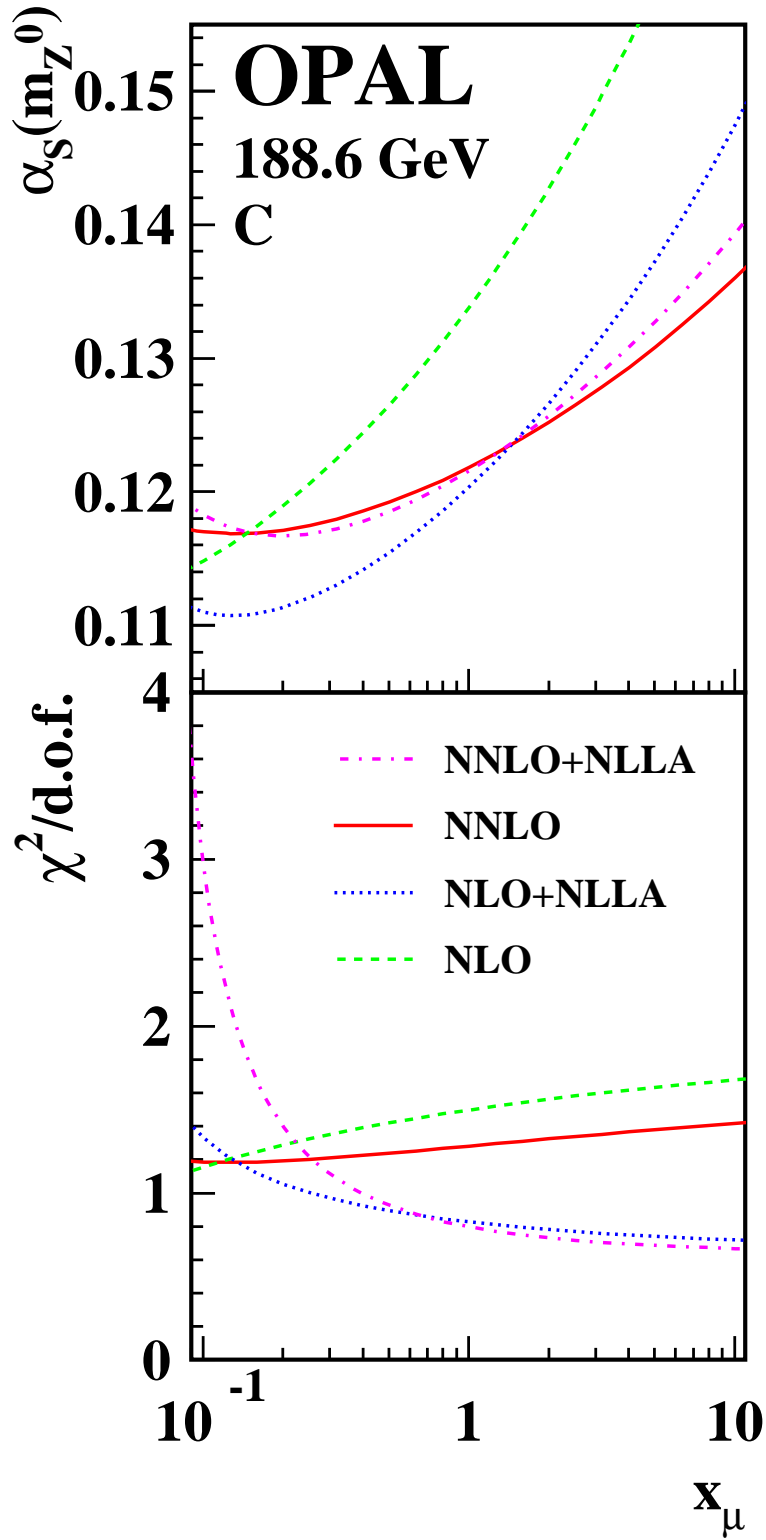}  % asvsxmu75.eps
\caption{Results for \as\ and \chisqd\ for fits of QCD predictions to the C distribution, as a function of \xmu.}
\label{FigureScaleDep}
\end{center}
\end{figure}

To assess the dependence of \as\ on the choice of the
renormalization scale, the fits 
to distributions of the six event shape variables
are repeated using NNLO, NNLO+NLLA, NLO
and NLO+NLLA predictions with $0.1<\xmu<10$.  
As an example, \as\ and the \chisqd\ for the \cp\ variable 
at $\roots=91$~GeV 
are shown as a function of \xmu\ in Fig.~\ref{FigureScaleDep}.  
The smallest \chisqd\ values for the fixed order
calculations arise at the smallest scales, while with the NLLA calculations
smaller \chisqd\ values occur nearer the physical scale \rs, i.e. \xmu=1.
Near this scale, smaller \chisqd\ values
are observed with the NNLO curves than with the respective NLO curves.

The NLO calculation yields a larger value of \as\ 
than 
the other calculations for $\xmu>0.2$.
The \asmz\ values using 
the NLO+NLLA, NNLO and NNLO+NLLA calculations cross near the natural\footnote{The terms inducing
the renormalization scale dependence resemble terms of higher order in \as\, weighted
with $\ln\xmu$.} choice of
the renormalization scale $\xmu=1$.  
The NLLA terms at $\xmu=1$ averaged over
the fit range are
almost identical to the \oaaa-terms in the NNLO calculation.  A
similar behaviour can be observed for \thr\ and \bt.

\section{Summary and conclusions}
In this paper we present 
determinations
of the strong coupling \as\
using event shape observable distributions at c.m.\ energies
between 91 and 209~GeV.  
Fits using NNLO and combined NNLO+NLLA predictions 
are used to extract \as.
Combining the results from the NNLO
fits to the six event shape observables \thr, \mh, \bw, \bt, \cp\ and
\ytwothree\ at the thirteen OPAL energy points 
yields 
\begin{eqnarray*}
  \asmz&=&\resultnnloA\resultnnloB\,,
\end{eqnarray*}
with a total uncertainty on \asmz\ of 2.6\%.  
Combining the results from the NNLO+NLLA fits yields
\begin{eqnarray*}
  \asmz&=&\resultnnlonllaA\resultnnlonllaB\,,
\end{eqnarray*}
 with a total uncertainty of 3.4\%.  
This result supersedes that presented in~\cite{OPALPR404} because
it is based on more complete theoretical predictions while
using the same experimental data and procedures.
The variations between the combined results for \as\ at the 
different energies are
consistent with the running of \as\ as predicted by QCD
and exclude the absence of running.
The \as\ results we find for all 13 energy points are given in
Tabs.~\ref{detailedasresultsnnlo}-\ref{detailedasresultsnlonlla}, 
for the NNLO, NNLO+NLLA, and NLO+NLLA calculations.

The investigation of the
renormalization scale dependence of \asmz\ shows a reduced dependence
on \xmu\ when NNLO or NNLO+NLLA predictions are used, compared to
analyses with NLO or NLO+NLLA predictions.  The more complete NNLO or
NNLO+NLLA QCD predictions thus lead to smaller theoretical
uncertainties in our analysis.  
Adding the NLLA terms to the NNLO predictions significantly improves the description of the data.
The standard procedure to quantify theoretical
uncertainties leads, however, to somewhat larger uncertainties in most
of the NNLO+NLLA fits.

{\small
\section*{Acknowledgements}
\par

This research was supported by the DFG cluster of excellence `Origin
and Structure of the Universe'.

We would like to thank A.~Gehrmann-De Ridder, T.~Gehr\-mann, E.~W.~N.~Glover and G.~Heinrich  
for providing the event shape coefficient distributions.

We particularly wish to thank the SL Division for the efficient operation
of the LEP accelerator at all energies
 and for their close cooperation with
our experimental group.  In addition to the support staff at our own
institutions we are pleased to acknowledge the  \\
Department of Energy, USA, \\
National Science Foundation, USA, \\
Particle Physics and Astronomy Research Council, UK, \\
Natural Sciences and Engineering Research Council, Canada, \\
Israel Science Foundation, administered by the Israel
Academy of Science and Humanities, \\
Benoziyo Center for High Energy Physics,\\
Japanese Ministry of Education, Culture, Sports, Science and
Technology (MEXT) and a grant under the MEXT International
Science Research Program,\\
Japanese Society for the Promotion of Science (JSPS),\\
German Israeli Bi-national Science Foundation (GIF), \\
Bundesministerium f\"ur Bildung und Forschung, Germany, \\
National Research Council of Canada, \\
Hungarian Foundation for Scientific Research, OTKA T-038240, 
and T-042864,\\
The NWO/NATO Fund for Scientific Research, the Netherlands.\\
}

\bibliographystyle{ephja}
\bibliography{papers}

%Tables%
% was TableNumberEvents.tex:
\begin{table}[h]
\caption{Year of data collection, energy range, mean c.m.\ energy, integrated luminosity ${\mathcal L}$
and numbers of selected events for each 
\Opal\ data sample used in this analysis, see also~\cite{OPALPR404}.
The horizontal lines  
divide the data into four energy ranges 
used for presentation purposes.}
\label{TableNumberEvents}
\begin{center}  
\begin{tabular}{ l r@{---}l l l r }
\hline\noalign{\smallskip}
Year       & \multicolumn{2}{l}{Range of \rs   }  & Mean $\rs$   & ${\cal L}$    & Selected \\
           & \multicolumn{2}{l}{[GeV]}  & [GeV]        & [pb$^{-1}$]   & events   \\
\noalign{\smallskip}\hline\noalign{\smallskip}
1996, 2000 &  91.0 & 91.5 &  91.3 &  14.7 & 395695 \\ 
\hline
1995, 1997 & 129.9 & 130.2 & 130.1 &   5.31 & 318   \\ 
1995, 1997 & 136.0 & 136.3 & 136.1 &   5.95 & 312   \\ 
\hline
1996       & 161.2 & 161.6 & 161.3 & 10.06 & 281    \\ 
1996       & 170.2 & 172.5 & 172.1 & 10.38 & 218    \\ 
1997       & 180.8 & 184.2 & 182.7 & 57.72 & 1077   \\ 
\hline
1998       & 188.3 & 189.1 & 188.6 & 185.2 & 3086   \\ 
1999       & 191.4 & 192.1 & 191.6 & 29.53 & 514    \\ 
1999       & 195.4 & 196.1 & 195.5 & 76.67 & 1137   \\ 
1999, 2000 & 199.1 & 200.2 & 199.5 & 79.27 & 1090   \\ 
1999, 2000 & 201.3 & 202.1 & 201.6 & 37.75 & 519    \\ 
2000       & 202.5 & 205.5 & 204.9 & 82.01 & 1130   \\ 
2000       & 205.5 & 208.9 & 206.6 & 138.8 & 1717   \\ 
\hline
\end{tabular}
\end{center}     
\end{table}
% was fitranges.tex:
\begin{table}[htb!]
\caption{Fit ranges at all c.m.\ energies.}  
\label{fitranges}
\begin{center}  
\begin{tabular}{ ccc }
\hline\noalign{\smallskip}
\thr & \mh & \bt \\
0.05--0.30 & 0.17--0.45 & 0.075--0.25 \\
\noalign{\smallskip}\hline\noalign{\smallskip}
\bw & \cp & \ytwothree \\
0.05--0.20 & 0.18-0.60 & 0.012--0.13 \\
\noalign{\smallskip}\hline
\end{tabular}  
\end{center}     
\end{table}
% was resultnnlo+-nlla.tex:
\begin{table*}[htb!]
\caption{Measurements of \as\ using 
NNLO predictions and
event shape distributions in four ranges of c.m.\ energy:
at 91.3~GeV, 133.1~GeV, 161--183~GeV (177.4~GeV on average) and
189--209~GeV (197.0~GeV on average).
}
\label{asresultsnnlo}
\begin{center}  
{\begin{tabular}{ ccrrrrrr }
\hline\noalign{\smallskip}
\roots\ [GeV] & Obs. & $\as(\roots)$ & $\pm$stat. & $\pm$exp. & $\pm$had. & $\pm$theo. & $\pm$tot. \\
\noalign{\smallskip}\hline\noalign{\smallskip}
% was asopal4_nnlo.tex:
 91.3&\thr&      0.1220& 0.0002& 0.0011& 0.0030& 0.0042 & 0.0053 \\
 91.3&\mh&       0.1228& 0.0002& 0.0008& 0.0026& 0.0028 & 0.0039 \\
 91.3&\bt&       0.1193& 0.0002& 0.0007& 0.0033& 0.0039 & 0.0052 \\
 91.3&\bw&       0.1201& 0.0001& 0.0014& 0.0010& 0.0021 & 0.0027 \\
 91.3&\cp&       0.1188& 0.0002& 0.0009& 0.0032& 0.0035 & 0.0048 \\
 91.3&\ytwothree&0.1202& 0.0002& 0.0025& 0.0005& 0.0019 & 0.0032 \betweenline
133.1&\thr&      0.1126& 0.0043& 0.0038& 0.0026& 0.0032 & 0.0071 \\
133.1&\mh&       0.1110& 0.0043& 0.0033& 0.0003& 0.0020 & 0.0058 \\
133.1&\bt&       0.1065& 0.0038& 0.0048& 0.0024& 0.0027 & 0.0071 \\
133.1&\bw&       0.1123& 0.0040& 0.0026& 0.0010& 0.0017 & 0.0052 \\
133.1&\cp&       0.1051& 0.0046& 0.0029& 0.0033& 0.0024 & 0.0068 \\
133.1&\ytwothree&0.1071& 0.0056& 0.0053& 0.0014& 0.0012 & 0.0079 \betweenline
177.4&\thr&      0.1088& 0.0029& 0.0027& 0.0013& 0.0028 & 0.0050 \\
177.4&\mh&       0.1081& 0.0028& 0.0042& 0.0011& 0.0018 & 0.0055 \\
177.4&\bt&       0.1051& 0.0023& 0.0031& 0.0024& 0.0026 & 0.0052 \\
177.4&\bw&       0.1047& 0.0024& 0.0029& 0.0013& 0.0012 & 0.0042 \\
177.4&\cp&       0.1067& 0.0028& 0.0030& 0.0015& 0.0025 & 0.0050 \\
177.4&\ytwothree&0.1084& 0.0041& 0.0031& 0.0005& 0.0013 & 0.0053 \betweenline
197.0&\thr&      0.1109& 0.0012& 0.0019& 0.0011& 0.0030 & 0.0039 \\
197.0&\mh&       0.1075& 0.0012& 0.0021& 0.0016& 0.0018 & 0.0034 \\
197.0&\bt&       0.1092& 0.0011& 0.0019& 0.0013& 0.0029 & 0.0039 \\
197.0&\bw&       0.1069& 0.0011& 0.0011& 0.0004& 0.0013 & 0.0021 \\
197.0&\cp&       0.1086& 0.0013& 0.0017& 0.0016& 0.0026 & 0.0037 \\
197.0&\ytwothree&0.1073& 0.0019& 0.0023& 0.0003& 0.0013 & 0.0033 \\ \noalign{\smallskip}\hline
\end{tabular}}
\end{center}  
\end{table*}

\clearpage

\begin{table*}[htb!]
\caption{Measurements of \as\ using 
NNLO predictions and
event shape distributions at 91.3~GeV. The employed covariance matrix includes
an estimate of the experimental systematic covariance.
The statistical error is taken from the fit with
only the statistical covariance.  
The experimental systematic 
uncertainty is given by the fit error with quadratically subtracted statistical uncertainty.
}
\label{NNLOminOvl}
\begin{center}  
{\begin{tabular}{ ccrrrrr }
\hline\noalign{\smallskip}
Obs. & $\as(91.3~{\rm GeV})$ & $\pm$stat. & $\pm$exp. & \chisqd \\
\noalign{\smallskip}\hline\noalign{\smallskip}
% was asopal4nnloMinovl.tex:
  %aus asopal_nnlo.TEX:          
\thr       & 0.1168 & 0.0002 & 0.0007  &  111/5 \\
\mh        & 0.1291 & 0.0002 & 0.0027  &  18.3/4 \\
\bt        & 0.1247 & 0.0002 & 0.0011  &   311/5 \\
\bw        & 0.1107 & 0.0001 & 0.0009  &   142/4 \\
\cp        & 0.1177 & 0.0002 & 0.0012  &  45.0/4 \\
\ytwothree & 0.1183 & 0.0002 & 0.0010  &   1.6/3 \\ \noalign{\smallskip} \hline
\end{tabular}}
\end{center} 
\end{table*}

\begin{table*}[htb!]
\caption{Measurements of \as\ using 
NNLO+NLLA predictions and
event shape distributions in four ranges of c.m.\ energy:
at 91.3~GeV, 133.1~GeV, 161--183~GeV (177.4~GeV on average) and
189--209~GeV (197.0~GeV on average).
}
\label{asresultsnnlonlla}
\begin{center}  
{\begin{tabular}{ ccrrrrrr }
\hline\noalign{\smallskip}
\roots\ [GeV] & Obs. & $\as(\roots)$ & $\pm$stat. & $\pm$exp. & $\pm$had. & $\pm$theo. & $\pm$tot. \\
\noalign{\smallskip}\hline\noalign{\smallskip}
% was asopal4_nnlo+nlla.tex:
 91.3&\thr      & 0.1219& 0.0002& 0.0012& 0.0030& 0.0041 & 0.0052 \\
 91.3&\mh       & 0.1207& 0.0002& 0.0008& 0.0022& 0.0033 & 0.0041 \\
 91.3&\bt       & 0.1213& 0.0002& 0.0010& 0.0023& 0.0048 & 0.0054 \\
 91.3&\bw       & 0.1164& 0.0001& 0.0013& 0.0011& 0.0041 & 0.0044 \\
 91.3&\cp       & 0.1186& 0.0002& 0.0009& 0.0030& 0.0046 & 0.0056 \\
 91.3&\ytwothree& 0.1195& 0.0002& 0.0025& 0.0004& 0.0023 & 0.0034 \betweenline
133.1&\thr      & 0.1128& 0.0044& 0.0040& 0.0025& 0.0032 & 0.0072 \\
133.1&\mh       & 0.1094& 0.0042& 0.0029& 0.0003& 0.0023 & 0.0056 \\
133.1&\bt       & 0.1085& 0.0041& 0.0055& 0.0016& 0.0034 & 0.0078 \\
133.1&\bw       & 0.1082& 0.0035& 0.0028& 0.0013& 0.0033 & 0.0057 \\
133.1&\cp       & 0.1049& 0.0047& 0.0034& 0.0027& 0.0030 & 0.0071 \\
133.1&\ytwothree& 0.1068& 0.0056& 0.0051& 0.0012& 0.0017 & 0.0079 \betweenline
177.4&\thr      & 0.1089& 0.0029& 0.0028& 0.0012& 0.0027 & 0.0050 \\
177.4&\mh       & 0.1069& 0.0027& 0.0039& 0.0013& 0.0021 & 0.0053 \\
177.4&\bt       & 0.1070& 0.0026& 0.0034& 0.0019& 0.0032 & 0.0057 \\
177.4&\bw       & 0.1018& 0.0023& 0.0026& 0.0012& 0.0027 & 0.0046 \\
177.4&\cp       & 0.1063& 0.0029& 0.0031& 0.0014& 0.0030 & 0.0054 \\
177.4&\ytwothree& 0.1079& 0.0040& 0.0031& 0.0005& 0.0016 & 0.0053 \betweenline
197.0&\thr      & 0.1110& 0.0013& 0.0020& 0.0012& 0.0029 & 0.0039 \\
197.0&\mh       & 0.1064& 0.0012& 0.0021& 0.0017& 0.0021 & 0.0036 \\
197.0&\bt       & 0.1112& 0.0012& 0.0021& 0.0011& 0.0035 & 0.0044 \\
197.0&\bw       & 0.1045& 0.0010& 0.0013& 0.0004& 0.0029 & 0.0034 \\
197.0&\cp       & 0.1083& 0.0013& 0.0019& 0.0017& 0.0032 & 0.0043 \\
197.0&\ytwothree& 0.1068& 0.0019& 0.0022& 0.0003& 0.0016 & 0.0033 \\\noalign{\smallskip}\hline
\end{tabular}}
\end{center}  
\end{table*}

\clearpage

\begin{table*}[htb!]
\caption{Measurements of \as\ using 
NNLO+NLLA predictions and
event shape distributions at 91.3~GeV. The employed covariance matrix includes
an estimate of the experimental systematic covariance.
The statistical error is taken from the fit with
only the statistical covariance.  
The experimental systematic 
uncertainty is given by the fit error with quadratically subtracted statistical uncertainty.
}
\label{NNLONLLAminOvl}
\begin{center}  
{\begin{tabular}{ ccrrrrr }
\hline\noalign{\smallskip}
Obs. & $\as(91.3~{\rm GeV})$ & $\pm$stat. & $\pm$exp. & \chisqd \\
\noalign{\smallskip}\hline\noalign{\smallskip}
% was asopal4nnlo+nllaMinovl.tex:
  %aus asopal_nnlo+nlla.TEX:          
\thr       & 0.1216 & 0.0002 & 0.0008  &  17.0/5 \\
\mh        & 0.1215 & 0.0002 & 0.0021  &   2.6/4 \\
\bt        & 0.1224 & 0.0002 & 0.0009  &   6.1/5 \\
\bw        & 0.1145 & 0.0001 & 0.0010  &   3.4/4 \\
\cp        & 0.1183 & 0.0002 & 0.0012  &   4.7/4 \\
\ytwothree & 0.1191 & 0.0002 & 0.0011  &   0.3/3 \\ \noalign{\smallskip} \hline
\end{tabular}}
\end{center} 
\end{table*}
% was asecmcomb+-all.tex:
\begin{table}[htb!]
\caption{Combined values of $\as(\roots)$ at the OPAL c.m.\ energy ranges
from NNLO (upper section) and NNLO+NLLA (lower section) analyses
together with the statistical, experimental, hadronisation, theory and total
uncertainties.}
\label{asecmcomb}
\begin{center}  
\begin{tabular}{ crrrrrrr }
\hline\noalign{\smallskip}
\roots\ [GeV] & $\as(\roots)$ & $\pm$stat. & $\pm$exp. & $\pm$had. & $\pm$theo. & $\pm$tot. \\
\noalign{\smallskip}\hline\noalign{\smallskip}
 \multicolumn{7}{c}{NNLO}\\
\hline
  % was asopalcomb_nnlo.tex :
 91.3 & 0.1205 & 0.0001 & 0.0011 & 0.0016 & 0.0027 & 0.0033 \\
133.1 & 0.1109 & 0.0036 & 0.0027 & 0.0011 & 0.0022 & 0.0051 \\
177.4 & 0.1057 & 0.0022 & 0.0027 & 0.0013 & 0.0020 & 0.0042 \\
197.0 & 0.1077 & 0.0010 & 0.0012 & 0.0006 & 0.0018 & 0.0025 \\
\noalign{\smallskip}\hline\noalign{\smallskip}
 \multicolumn{7}{c}{NNLO+NLLA}\\
\hline
  % was asopalcomb_nnlo+nlla.tex :
 91.3 & 0.1196 & 0.0002 & 0.0012 & 0.0013 & 0.0036 & 0.0040 \\
133.1 & 0.1088 & 0.0036 & 0.0029 & 0.0009 & 0.0030 & 0.0056 \\
177.4 & 0.1042 & 0.0024 & 0.0027 & 0.0009 & 0.0027 & 0.0046 \\
197.0 & 0.1069 & 0.0010 & 0.0016 & 0.0008 & 0.0029 & 0.0036 \\
  \noalign{\smallskip}\hline
\end{tabular}
\end{center}  
\end{table}

\begin{table}[htb!]
\caption{Combined values of \asmz\ at OPAL c.m.\ energy ranges
from NNLO (upper section) and NNLO+NLLA (lower section) analyses
together with the statistical, experimental, hadronisation, theory and total
uncertainties.}
\label{asecmcomb_all}
\hspace{-.7cm}
\begin{center}  
\begin{tabular}{  r@{---}l r r r r r r }
\hline\noalign{\smallskip}
 \multicolumn{2}{c}{\roots [GeV]} & \asmz & $\pm$stat. & $\pm$exp. & $\pm$had. & $\pm$theo. & $\pm$tot. \\
\noalign{\smallskip}\hline\noalign{\smallskip}
\multicolumn{8}{c}{NNLO}\\
\hline
  % was asopalcomb_alltab_nnlo:wo91.tex :
 130.1 & 206.6 &   0.1200 &   0.0012 &   0.0016 &   0.0008 &   0.0023 & 0.0032\\
  % was asopalcomb_alltab_nnlo.tex :
 91.3 & 206.6 & 0.1201 & 0.0008 & 0.0013 & 0.0010 & 0.0024 & 0.0031\\
\noalign{\smallskip}\hline\noalign{\smallskip}
 \multicolumn{8}{c}{NNLO+NLLA}\\
\hline
  % was asopalcomb_alltab_nnlo+nlla:wo91.tex :
 130.1 & 206.6 &   0.1186 &   0.0011 &   0.0020 &   0.0009 &   0.0036 & 0.0040\\
  % was asopalcomb_alltab_nnlo+nlla.tex :
 91.3 & 206.6 &  0.1189 &  0.0008 &  0.0016 &  0.0010 &  0.0036 & 0.0041\\
\noalign{\smallskip}\hline
\end{tabular}
\end{center}  
\end{table}
 % was asvarcomb.tex :
\begin{table}[htb!]
\caption{Combined values of \asmz\ for each observable from NNLO
(upper section) and NNLO+NLLA (lower section) analyses together with
the statistical, experimental, hadronisation, theory and total uncertainties.}
\label{asvarcomb}
\begin{center}
\begin{tabular}{ crrrrrr }
\hline\noalign{\smallskip}
Obs. & \asmz & $\pm$stat. & $\pm$exp. & $\pm$had. & $\pm$theo. & $\pm$tot. \\
\noalign{\smallskip}\hline\noalign{\smallskip}
 \multicolumn{7}{c}{NNLO}\\
\hline
  % was asopalcombv_nnlo.tex :
\thr & 0.1230 & 0.0010 & 0.0019 & 0.0018 & 0.0039 & 0.0048 \\
\mh  & 0.1212 & 0.0009 & 0.0017 & 0.0020 & 0.0025 & 0.0037 \\
\bt  & 0.1205 & 0.0009 & 0.0018 & 0.0016 & 0.0037 & 0.0045 \\
\bw  & 0.1195 & 0.0009 & 0.0012 & 0.0007 & 0.0018 & 0.0024 \\
\cp  & 0.1199 & 0.0011 & 0.0016 & 0.0023 & 0.0033 & 0.0045 \\
\ytwothree & 0.1199 & 0.0008 & 0.0026 & 0.0004 & 0.0018 & 0.0033 \\
  \noalign{\smallskip}\hline\noalign{\smallskip}
 \multicolumn{7}{c}{NNLO+NLLA}\\
\hline
  % was asopalcombv_nnlo+nlla.tex :
\thr & 0.1230 & 0.0010 & 0.0020 & 0.0020 & 0.0038 & 0.0048 \\
\mh  & 0.1194 & 0.0009 & 0.0017 & 0.0020 & 0.0029 & 0.0040 \\
\bt  & 0.1227 & 0.0009 & 0.0020 & 0.0016 & 0.0045 & 0.0053 \\
\bw  & 0.1159 & 0.0009 & 0.0015 & 0.0007 & 0.0037 & 0.0042 \\
\cp  & 0.1196 & 0.0011 & 0.0019 & 0.0023 & 0.0042 & 0.0053 \\
\ytwothree & 0.1192 & 0.0010 & 0.0026 & 0.0004 & 0.0021 & 0.0035 \\
 \noalign{\smallskip}\hline
\end{tabular}
\end{center}
\end{table}

%Figures%
% was McVsCalc.tex :
% was esdistributions123.tex :
% was asplotopal,asvscme.tex :
% was asobsscatter.tex :
% was FigureScaleDep.tex :
\clearpage
\begin{appendix}
\section{Appendix: Detailed results}\label{separateFits}
\begin{table*}[htb!]
\caption{Results of NNLO fits to event shape observable distributions
at the OPAL c.m.\ energies.
The \chisqd\ values are based on the statistical errors only.}
\label{detailedasresultsnnlo}
\begin{center}  
\scalebox{0.9}
{\begin{tabular}{ ccrrrrrc }
\hline\noalign{\smallskip}
\roots\ [GeV] & Obs. & $\as(\roots)$ & $\pm$stat. & $\pm$exp. & $\pm$had. & $\pm$theo. & \chisqd \\
\noalign{\smallskip}\hline\noalign{\smallskip}
% was asopal_nnlo_1.tex :
91.3 & \thr &  0.1220 &  0.0002 &  0.0011 &  0.0030 &  0.0042 & $440.1/ 5$ \\
91.3 & \mh &  0.1228 &  0.0002 &  0.0008 &  0.0027 &  0.0028 & $393.8/ 4$ \\
91.3 & \bt &  0.1193 &  0.0001 &  0.0007 &  0.0033 &  0.0039 & $1142.4/ 5$ \\
91.3 & \bw &  0.1201 &  0.0001 &  0.0013 &  0.0010 &  0.0021 & $446.6/ 4$ \\
91.3 & \cp &  0.1188 &  0.0002 &  0.0009 &  0.0032 &  0.0035 & $531.8/ 4$ \\
91.3 & \ytwothree &  0.1202 &  0.0002 &  0.0025 &  0.0005 &  0.0019 & $31.5/ 3$ \\
\hline
130.1 & \thr &  0.1179 &  0.0057 &  0.0030 &  0.0025 &  0.0037 & $ 8.3/ 5$ \\
130.1 & \mh &  0.1158 &  0.0057 &  0.0027 &  0.0004 &  0.0023 & $ 6.5/ 4$ \\
130.1 & \bt &  0.1098 &  0.0051 &  0.0045 &  0.0019 &  0.0029 & $ 8.2/ 5$ \\
130.1 & \bw &  0.1155 &  0.0049 &  0.0020 &  0.0006 &  0.0018 & $ 0.8/ 4$ \\
130.1 & \cp &  0.1106 &  0.0061 &  0.0019 &  0.0028 &  0.0027 & $ 7.8/ 4$ \\
130.1 & \ytwothree &  0.1106 &  0.0075 &  0.0046 &  0.0016 &  0.0014 & $ 4.7/ 3$ \\
\hline
136.1 & \thr &  0.1019 &  0.0062 &  0.0080 &  0.0029 &  0.0022 & $12.0/ 5$ \\
136.1 & \mh &  0.1012 &  0.0061 &  0.0067 &  0.0016 &  0.0014 & $ 5.1/ 4$ \\
136.1 & \bt &  0.1002 &  0.0053 &  0.0072 &  0.0034 &  0.0023 & $ 8.7/ 5$ \\
136.1 & \bw &  0.1021 &  0.0051 &  0.0082 &  0.0024 &  0.0013 & $ 7.0/ 4$ \\
136.1 & \cp &  0.0932 &  0.0065 &  0.0073 &  0.0045 &  0.0016 & $10.6/ 4$ \\
136.1 & \ytwothree &  0.0999 &  0.0076 &  0.0095 &  0.0009 &  0.0010 & $ 3.0/ 3$ \\
\hline
161.3 & \thr &  0.1068 &  0.0065 &  0.0031 &  0.0025 &  0.0026 & $ 6.4/ 5$ \\
161.3 & \mh &  0.1073 &  0.0064 &  0.0040 &  0.0009 &  0.0017 & $ 3.9/ 4$ \\
161.3 & \bt &  0.1001 &  0.0055 &  0.0040 &  0.0034 &  0.0022 & $ 8.7/ 5$ \\
161.3 & \bw &  0.1022 &  0.0054 &  0.0023 &  0.0034 &  0.0012 & $ 9.4/ 4$ \\
161.3 & \cp &  0.1019 &  0.0066 &  0.0044 &  0.0034 &  0.0021 & $ 8.2/ 4$ \\
161.3 & \ytwothree &  0.1060 &  0.0086 &  0.0020 &  0.0008 &  0.0013 & $ 2.0/ 3$ \\
\hline
172.1 & \thr &  0.1087 &  0.0076 &  0.0072 &  0.0030 &  0.0029 & $ 6.3/ 5$ \\
172.1 & \mh &  0.1068 &  0.0073 &  0.0062 &  0.0009 &  0.0018 & $ 5.0/ 4$ \\
172.1 & \bt &  0.1018 &  0.0064 &  0.0047 &  0.0021 &  0.0023 & $ 4.5/ 5$ \\
172.1 & \bw &  0.1001 &  0.0064 &  0.0039 &  0.0007 &  0.0011 & $ 1.9/ 4$ \\
172.1 & \cp &  0.1046 &  0.0075 &  0.0071 &  0.0033 &  0.0024 & $ 7.0/ 4$ \\
172.1 & \ytwothree &  0.1044 &  0.0104 &  0.0132 &  0.0013 &  0.0012 & $ 2.6/ 3$ \\
\hline
182.7 & \thr &  0.1098 &  0.0035 &  0.0025 &  0.0018 &  0.0029 & $ 0.6/ 5$ \\
182.7 & \mh &  0.1091 &  0.0033 &  0.0044 &  0.0015 &  0.0018 & $ 4.4/ 4$ \\
182.7 & \bt &  0.1078 &  0.0028 &  0.0029 &  0.0021 &  0.0028 & $ 9.0/ 5$ \\
182.7 & \bw &  0.1071 &  0.0029 &  0.0032 &  0.0005 &  0.0013 & $ 4.3/ 4$ \\
182.7 & \cp &  0.1086 &  0.0034 &  0.0026 &  0.0018 &  0.0026 & $ 1.1/ 4$ \\
182.7 & \ytwothree &  0.1103 &  0.0048 &  0.0040 &  0.0003 &  0.0014 & $ 1.4/ 3$ \\
\hline
                                  
\end{tabular}}
\end{center}  
\end{table*}

\begin{table*}[htb!]
\caption{Results of NNLO fits to event shape observable distributions
at the OPAL c.m.\ energies.
The \chisqd\ values are based on the statistical errors only.}
\label{detailedasresultsnnlo2}
\begin{center}  
\scalebox{0.9}
{\begin{tabular}{ ccrrrrrc }
\hline\noalign{\smallskip}
\roots\ [GeV] & Obs. & $\as(\roots)$ & $\pm$stat. & $\pm$exp. & $\pm$had. & $\pm$theo. & \chisqd \\
\noalign{\smallskip}\hline\noalign{\smallskip}
% was asopal_nnlo_2.tex :
188.6 & \thr &  0.1133 &  0.0020 &  0.0015 &  0.0014 &  0.0032 & $14.3/ 5$ \\
188.6 & \mh &  0.1090 &  0.0020 &  0.0021 &  0.0018 &  0.0018 & $ 4.9/ 4$ \\
188.6 & \bt &  0.1100 &  0.0017 &  0.0013 &  0.0011 &  0.0030 & $ 5.0/ 5$ \\
188.6 & \bw &  0.1074 &  0.0017 &  0.0006 &  0.0006 &  0.0014 & $ 2.1/ 4$ \\
188.6 & \cp &  0.1094 &  0.0021 &  0.0012 &  0.0016 &  0.0027 & $ 6.4/ 4$ \\
188.6 & \ytwothree &  0.1083 &  0.0029 &  0.0020 &  0.0004 &  0.0013 & $ 3.5/ 3$ \\
\hline
191.6 & \thr &  0.1112 &  0.0050 &  0.0085 &  0.0017 &  0.0031 & $ 1.9/ 5$ \\
191.6 & \mh &  0.1094 &  0.0048 &  0.0055 &  0.0014 &  0.0019 & $ 4.1/ 4$ \\
191.6 & \bt &  0.1015 &  0.0046 &  0.0089 &  0.0025 &  0.0022 & $10.2/ 5$ \\
191.6 & \bw &  0.1044 &  0.0042 &  0.0050 &  0.0005 &  0.0013 & $ 0.8/ 4$ \\
191.6 & \cp &  0.1053 &  0.0053 &  0.0060 &  0.0020 &  0.0024 & $ 0.3/ 4$ \\
191.6 & \ytwothree &  0.1051 &  0.0071 &  0.0066 &  0.0005 &  0.0012 & $ 0.0/ 3$ \\
\hline
195.5 & \thr &  0.1119 &  0.0033 &  0.0040 &  0.0021 &  0.0031 & $ 9.6/ 5$ \\
195.5 & \mh &  0.1057 &  0.0033 &  0.0032 &  0.0018 &  0.0016 & $ 2.0/ 4$ \\
195.5 & \bt &  0.1081 &  0.0029 &  0.0026 &  0.0016 &  0.0028 & $ 6.7/ 5$ \\
195.5 & \bw &  0.1062 &  0.0029 &  0.0018 &  0.0007 &  0.0013 & $ 4.5/ 4$ \\
195.5 & \cp &  0.1087 &  0.0035 &  0.0035 &  0.0017 &  0.0026 & $ 1.9/ 4$ \\
195.5 & \ytwothree &  0.1008 &  0.0048 &  0.0046 &  0.0004 &  0.0010 & $ 1.4/ 3$ \\
\hline
199.5 & \thr &  0.1105 &  0.0034 &  0.0057 &  0.0021 &  0.0029 & $ 8.5/ 5$ \\
199.5 & \mh &  0.1061 &  0.0034 &  0.0028 &  0.0007 &  0.0017 & $ 7.1/ 4$ \\
199.5 & \bt &  0.1102 &  0.0030 &  0.0039 &  0.0023 &  0.0030 & $ 8.5/ 5$ \\
199.5 & \bw &  0.1076 &  0.0029 &  0.0020 &  0.0010 &  0.0013 & $ 3.9/ 4$ \\
199.5 & \cp &  0.1077 &  0.0036 &  0.0039 &  0.0012 &  0.0026 & $ 5.5/ 4$ \\
199.5 & \ytwothree &  0.1130 &  0.0049 &  0.0045 &  0.0009 &  0.0015 & $ 4.8/ 3$ \\
\hline
201.6 & \thr &  0.1168 &  0.0048 &  0.0051 &  0.0013 &  0.0035 & $ 1.5/ 5$ \\
201.6 & \mh &  0.1045 &  0.0049 &  0.0047 &  0.0012 &  0.0016 & $12.2/ 4$ \\
201.6 & \bt &  0.1072 &  0.0044 &  0.0027 &  0.0015 &  0.0027 & $ 8.4/ 5$ \\
201.6 & \bw &  0.1057 &  0.0043 &  0.0026 &  0.0017 &  0.0012 & $ 7.7/ 4$ \\
201.6 & \cp &  0.1092 &  0.0052 &  0.0061 &  0.0024 &  0.0026 & $11.3/ 4$ \\
201.6 & \ytwothree &  0.1107 &  0.0071 &  0.0076 &  0.0008 &  0.0014 & $ 1.4/ 3$ \\
\hline
204.9 & \thr &  0.1111 &  0.0032 &  0.0032 &  0.0015 &  0.0031 & $ 8.7/ 5$ \\
204.9 & \mh &  0.1134 &  0.0029 &  0.0036 &  0.0006 &  0.0021 & $ 7.0/ 4$ \\
204.9 & \bt &  0.1091 &  0.0028 &  0.0057 &  0.0011 &  0.0029 & $ 4.7/ 5$ \\
204.9 & \bw &  0.1108 &  0.0027 &  0.0034 &  0.0005 &  0.0015 & $ 3.1/ 4$ \\
204.9 & \cp &  0.1085 &  0.0033 &  0.0068 &  0.0017 &  0.0027 & $ 3.3/ 4$ \\
204.9 & \ytwothree &  0.1090 &  0.0045 &  0.0060 &  0.0007 &  0.0013 & $ 3.7/ 3$ \\
\hline
206.6 & \thr &  0.1066 &  0.0026 &  0.0019 &  0.0018 &  0.0026 & $ 2.9/ 5$ \\
206.6 & \mh &  0.1055 &  0.0025 &  0.0017 &  0.0021 &  0.0016 & $ 0.8/ 4$ \\
206.6 & \bt &  0.1100 &  0.0021 &  0.0027 &  0.0010 &  0.0030 & $ 0.8/ 5$ \\
206.6 & \bw &  0.1061 &  0.0022 &  0.0024 &  0.0006 &  0.0013 & $ 0.9/ 4$ \\
206.6 & \cp &  0.1079 &  0.0026 &  0.0020 &  0.0017 &  0.0026 & $ 1.6/ 4$ \\
206.6 & \ytwothree &  0.1058 &  0.0037 &  0.0026 &  0.0003 &  0.0012 & $ 0.6/ 3$ \\
\hline
\end{tabular}}
\end{center} 
\end{table*}

\begin{table*}[htb!]
\caption{Results of NNLO+NLLA fits to event shape observable distributions
at the OPAL c.m.\ energies.
The \chisqd\ values are based on the statistical errors only.}
\label{detailedasresultsnnlonlla}
\begin{center} 
\scalebox{0.9}
{\begin{tabular}{ ccrrrrrc }
\hline\noalign{\smallskip}
\roots\ [GeV] & Obs. & $\as(\roots)$ & $\pm$stat. & $\pm$exp. & $\pm$had. & $\pm$theo. & \chisqd \\
\noalign{\smallskip}\hline\noalign{\smallskip}
% was asopal_nnlo+nlla_1.tex :
91.3 & \thr &  0.1219 &  0.0002 &  0.0012 &  0.0030 &  0.0041 & $120.5/ 5$ \\
91.3 & \mh &  0.1206 &  0.0002 &  0.0008 &  0.0022 &  0.0033 & $66.8/ 4$ \\
91.3 & \bt &  0.1213 &  0.0002 &  0.0009 &  0.0023 &  0.0048 & $40.0/ 5$ \\
91.3 & \bw &  0.1164 &  0.0001 &  0.0013 &  0.0011 &  0.0041 & $108.9/ 4$ \\
91.3 & \cp &  0.1187 &  0.0002 &  0.0009 &  0.0030 &  0.0046 & $51.2/ 4$ \\
91.3 & \ytwothree &  0.1195 &  0.0002 &  0.0025 &  0.0005 &  0.0023 & $ 8.0/ 3$ \\
\hline
130.1 & \thr &  0.1178 &  0.0058 &  0.0031 &  0.0025 &  0.0037 & $ 8.1/ 5$ \\
130.1 & \mh &  0.1138 &  0.0054 &  0.0022 &  0.0008 &  0.0026 & $ 6.7/ 4$ \\
130.1 & \bt &  0.1130 &  0.0056 &  0.0050 &  0.0024 &  0.0039 & $ 5.5/ 5$ \\
130.1 & \bw &  0.1119 &  0.0046 &  0.0019 &  0.0009 &  0.0036 & $ 1.0/ 4$ \\
130.1 & \cp &  0.1113 &  0.0063 &  0.0024 &  0.0019 &  0.0038 & $ 6.4/ 4$ \\
130.1 & \ytwothree &  0.1105 &  0.0075 &  0.0044 &  0.0015 &  0.0019 & $ 4.4/ 3$ \\
\hline
136.1 & \thr &  0.1026 &  0.0063 &  0.0084 &  0.0025 &  0.0023 & $11.0/ 5$ \\
136.1 & \mh &  0.1003 &  0.0059 &  0.0066 &  0.0018 &  0.0017 & $ 4.7/ 4$ \\
136.1 & \bt &  0.1012 &  0.0056 &  0.0079 &  0.0034 &  0.0026 & $ 8.2/ 5$ \\
136.1 & \bw &  0.0993 &  0.0047 &  0.0076 &  0.0024 &  0.0024 & $ 7.3/ 4$ \\
136.1 & \cp &  0.0927 &  0.0064 &  0.0073 &  0.0044 &  0.0016 & $10.5/ 4$ \\
136.1 & \ytwothree &  0.0996 &  0.0075 &  0.0093 &  0.0008 &  0.0012 & $ 2.9/ 3$ \\
\hline
161.3 & \thr &  0.1076 &  0.0066 &  0.0032 &  0.0020 &  0.0028 & $ 5.6/ 5$ \\
161.3 & \mh &  0.1068 &  0.0062 &  0.0037 &  0.0017 &  0.0022 & $ 3.1/ 4$ \\
161.3 & \bt &  0.1021 &  0.0059 &  0.0041 &  0.0029 &  0.0027 & $ 7.5/ 5$ \\
161.3 & \bw &  0.1001 &  0.0051 &  0.0022 &  0.0030 &  0.0025 & $ 9.1/ 4$ \\
161.3 & \cp &  0.1024 &  0.0067 &  0.0046 &  0.0029 &  0.0028 & $ 7.4/ 4$ \\
161.3 & \ytwothree &  0.1057 &  0.0085 &  0.0020 &  0.0007 &  0.0015 & $ 1.9/ 3$ \\
\hline
172.1 & \thr &  0.1080 &  0.0076 &  0.0071 &  0.0032 &  0.0024 & $ 6.7/ 5$ \\
172.1 & \mh &  0.1047 &  0.0069 &  0.0057 &  0.0015 &  0.0019 & $ 5.6/ 4$ \\
172.1 & \bt &  0.1038 &  0.0069 &  0.0054 &  0.0015 &  0.0029 & $ 3.7/ 5$ \\
172.1 & \bw &  0.0978 &  0.0060 &  0.0038 &  0.0007 &  0.0023 & $ 2.0/ 4$ \\
172.1 & \cp &  0.1038 &  0.0075 &  0.0072 &  0.0034 &  0.0025 & $ 7.2/ 4$ \\
172.1 & \ytwothree &  0.1038 &  0.0102 &  0.0129 &  0.0013 &  0.0015 & $ 2.6/ 3$ \\
\hline
182.7 & \thr &  0.1098 &  0.0035 &  0.0026 &  0.0018 &  0.0028 & $ 0.4/ 5$ \\
182.7 & \mh &  0.1080 &  0.0032 &  0.0040 &  0.0017 &  0.0022 & $ 3.7/ 4$ \\
182.7 & \bt &  0.1102 &  0.0031 &  0.0032 &  0.0015 &  0.0035 & $ 5.5/ 5$ \\
182.7 & \bw &  0.1047 &  0.0027 &  0.0030 &  0.0004 &  0.0029 & $ 3.5/ 4$ \\
182.7 & \cp &  0.1083 &  0.0035 &  0.0028 &  0.0019 &  0.0032 & $ 0.6/ 4$ \\
182.7 & \ytwothree &  0.1098 &  0.0047 &  0.0039 &  0.0003 &  0.0017 & $ 1.3/ 3$ \\
\hline
\end{tabular}}
\end{center}  
\end{table*}

\begin{table*}[htb!]
\caption{Results of NNLO+NLLA fits to event shape observable distributions
at the OPAL c.m.\ energies.
The \chisqd\ values are based on the statistical errors only.}
\begin{center}  
\scalebox{0.9}
{\begin{tabular}{ ccrrrrrc }
\hline\noalign{\smallskip}
\roots\ [GeV] & Obs. & $\as(\roots)$ & $\pm$stat. & $\pm$exp. & $\pm$had. & $\pm$theo. & \chisqd \\
\noalign{\smallskip}\hline\noalign{\smallskip}
% was asopal_nnlo+nlla_2.tex :
188.6 & \thr &  0.1133 &  0.0020 &  0.0016 &  0.0012 &  0.0032 & $13.6/ 5$ \\
188.6 & \mh &  0.1077 &  0.0019 &  0.0020 &  0.0019 &  0.0021 & $ 4.3/ 4$ \\
188.6 & \bt &  0.1119 &  0.0019 &  0.0014 &  0.0012 &  0.0036 & $ 2.1/ 5$ \\
188.6 & \bw &  0.1048 &  0.0016 &  0.0006 &  0.0006 &  0.0029 & $ 1.1/ 4$ \\
188.6 & \cp &  0.1092 &  0.0021 &  0.0014 &  0.0018 &  0.0033 & $ 4.0/ 4$ \\
188.6 & \ytwothree &  0.1077 &  0.0028 &  0.0020 &  0.0004 &  0.0016 & $ 3.6/ 3$ \\
\hline
191.6 & \thr &  0.1110 &  0.0051 &  0.0085 &  0.0016 &  0.0028 & $ 2.1/ 5$ \\
191.6 & \mh &  0.1076 &  0.0046 &  0.0055 &  0.0011 &  0.0021 & $ 4.8/ 4$ \\
191.6 & \bt &  0.1040 &  0.0050 &  0.0099 &  0.0021 &  0.0028 & $ 7.9/ 5$ \\
191.6 & \bw &  0.1019 &  0.0040 &  0.0049 &  0.0005 &  0.0026 & $ 0.9/ 4$ \\
191.6 & \cp &  0.1047 &  0.0053 &  0.0060 &  0.0020 &  0.0027 & $ 0.3/ 4$ \\
191.6 & \ytwothree &  0.1046 &  0.0070 &  0.0066 &  0.0005 &  0.0014 & $ 0.0/ 3$ \\
\hline
195.5 & \thr &  0.1121 &  0.0034 &  0.0041 &  0.0020 &  0.0031 & $ 9.0/ 5$ \\
195.5 & \mh &  0.1047 &  0.0032 &  0.0032 &  0.0020 &  0.0020 & $ 1.2/ 4$ \\
195.5 & \bt &  0.1104 &  0.0032 &  0.0028 &  0.0012 &  0.0035 & $ 3.9/ 5$ \\
195.5 & \bw &  0.1038 &  0.0027 &  0.0017 &  0.0005 &  0.0028 & $ 3.7/ 4$ \\
195.5 & \cp &  0.1083 &  0.0036 &  0.0034 &  0.0018 &  0.0031 & $ 1.6/ 4$ \\
195.5 & \ytwothree &  0.1005 &  0.0047 &  0.0046 &  0.0003 &  0.0013 & $ 1.2/ 3$ \\
\hline
199.5 & \thr &  0.1109 &  0.0035 &  0.0059 &  0.0017 &  0.0031 & $ 7.2/ 5$ \\
199.5 & \mh &  0.1052 &  0.0033 &  0.0029 &  0.0010 &  0.0020 & $ 6.0/ 4$ \\
199.5 & \bt &  0.1129 &  0.0034 &  0.0045 &  0.0019 &  0.0039 & $ 4.9/ 5$ \\
199.5 & \bw &  0.1053 &  0.0028 &  0.0020 &  0.0006 &  0.0029 & $ 2.5/ 4$ \\
199.5 & \cp &  0.1077 &  0.0037 &  0.0040 &  0.0013 &  0.0032 & $ 4.2/ 4$ \\
199.5 & \ytwothree &  0.1124 &  0.0048 &  0.0044 &  0.0009 &  0.0019 & $ 4.9/ 3$ \\
\hline
201.6 & \thr &  0.1172 &  0.0049 &  0.0053 &  0.0010 &  0.0037 & $ 1.0/ 5$ \\
201.6 & \mh &  0.1036 &  0.0048 &  0.0047 &  0.0010 &  0.0019 & $11.7/ 4$ \\
201.6 & \bt &  0.1097 &  0.0048 &  0.0036 &  0.0011 &  0.0034 & $ 6.5/ 5$ \\
201.6 & \bw &  0.1035 &  0.0041 &  0.0025 &  0.0013 &  0.0028 & $ 7.1/ 4$ \\
201.6 & \cp &  0.1104 &  0.0054 &  0.0068 &  0.0020 &  0.0038 & $ 9.3/ 4$ \\
201.6 & \ytwothree &  0.1104 &  0.0070 &  0.0074 &  0.0007 &  0.0018 & $ 1.2/ 3$ \\
\hline
204.9 & \thr &  0.1108 &  0.0032 &  0.0031 &  0.0014 &  0.0028 & $ 9.9/ 5$ \\
204.9 & \mh &  0.1117 &  0.0029 &  0.0034 &  0.0005 &  0.0024 & $ 7.9/ 4$ \\
204.9 & \bt &  0.1108 &  0.0030 &  0.0060 &  0.0011 &  0.0034 & $ 4.5/ 5$ \\
204.9 & \bw &  0.1077 &  0.0025 &  0.0032 &  0.0004 &  0.0032 & $ 4.7/ 4$ \\
204.9 & \cp &  0.1081 &  0.0033 &  0.0070 &  0.0017 &  0.0031 & $ 3.3/ 4$ \\
204.9 & \ytwothree &  0.1085 &  0.0044 &  0.0060 &  0.0007 &  0.0017 & $ 3.5/ 3$ \\
\hline
206.6 & \thr &  0.1068 &  0.0027 &  0.0019 &  0.0021 &  0.0025 & $ 2.1/ 5$ \\
206.6 & \mh &  0.1042 &  0.0025 &  0.0017 &  0.0022 &  0.0019 & $ 0.7/ 4$ \\
206.6 & \bt &  0.1116 &  0.0023 &  0.0030 &  0.0012 &  0.0035 & $ 1.5/ 5$ \\
206.6 & \bw &  0.1035 &  0.0021 &  0.0022 &  0.0005 &  0.0028 & $ 1.4/ 4$ \\
206.6 & \cp &  0.1076 &  0.0027 &  0.0021 &  0.0018 &  0.0031 & $ 0.9/ 4$ \\
206.6 & \ytwothree &  0.1054 &  0.0036 &  0.0025 &  0.0004 &  0.0015 & $ 0.5/ 3$ \\
\hline
\end{tabular}}
\end{center}  
\end{table*}

\begin{table*}[htb!]
\caption{Results of NLO+NLLA fits to event shape observable distributions
at the OPAL c.m.\ energies.
The \chisqd\ values are based on the statistical errors only.}
\begin{center}  
\scalebox{0.9}
{\begin{tabular}{ ccrrrrrc }
\hline\noalign{\smallskip}
\roots\ [GeV] & Obs. & $\as(\roots)$ & $\pm$stat. & $\pm$exp. & $\pm$had. & $\pm$theo. & \chisqd \\
\noalign{\smallskip}\hline\noalign{\smallskip}
% was asopal_nlo+nlla_1.tex :
91.3 & \thr &  0.1233 &  0.0002 &  0.0012 &  0.0031 &  0.0073 & $247.1/ 5$ \\
91.3 & \mh &  0.1196 &  0.0002 &  0.0008 &  0.0020 &  0.0057 & $36.9/ 4$ \\
91.3 & \bt &  0.1226 &  0.0002 &  0.0009 &  0.0029 &  0.0076 & $190.5/ 5$ \\
91.3 & \bw &  0.1145 &  0.0001 &  0.0013 &  0.0012 &  0.0055 & $140.2/ 4$ \\
91.3 & \cp &  0.1173 &  0.0002 &  0.0009 &  0.0031 &  0.0064 & $144.3/ 4$ \\
91.3 & \ytwothree &  0.1192 &  0.0002 &  0.0025 &  0.0005 &  0.0044 & $ 4.5/ 3$ \\
\hline
130.1 & \thr &  0.1187 &  0.0059 &  0.0032 &  0.0027 &  0.0065 & $ 8.6/ 5$ \\
130.1 & \mh &  0.1129 &  0.0054 &  0.0021 &  0.0008 &  0.0048 & $ 6.4/ 4$ \\
130.1 & \bt &  0.1138 &  0.0057 &  0.0050 &  0.0024 &  0.0063 & $ 5.8/ 5$ \\
130.1 & \bw &  0.1100 &  0.0046 &  0.0019 &  0.0010 &  0.0050 & $ 1.1/ 4$ \\
130.1 & \cp &  0.1100 &  0.0061 &  0.0023 &  0.0022 &  0.0054 & $ 6.8/ 4$ \\
130.1 & \ytwothree &  0.1104 &  0.0074 &  0.0043 &  0.0014 &  0.0035 & $ 4.2/ 3$ \\
\hline
136.1 & \thr &  0.1031 &  0.0064 &  0.0085 &  0.0027 &  0.0044 & $11.6/ 5$ \\
136.1 & \mh &  0.0995 &  0.0058 &  0.0066 &  0.0018 &  0.0032 & $ 4.4/ 4$ \\
136.1 & \bt &  0.1017 &  0.0057 &  0.0079 &  0.0036 &  0.0048 & $ 8.7/ 5$ \\
136.1 & \bw &  0.0976 &  0.0047 &  0.0076 &  0.0024 &  0.0036 & $ 7.3/ 4$ \\
136.1 & \cp &  0.0919 &  0.0063 &  0.0071 &  0.0044 &  0.0030 & $10.6/ 4$ \\
136.1 & \ytwothree &  0.0995 &  0.0075 &  0.0094 &  0.0008 &  0.0026 & $ 2.9/ 3$ \\
\hline
161.3 & \thr &  0.1083 &  0.0067 &  0.0032 &  0.0023 &  0.0051 & $ 5.9/ 5$ \\
161.3 & \mh &  0.1061 &  0.0062 &  0.0035 &  0.0019 &  0.0041 & $ 2.7/ 4$ \\
161.3 & \bt &  0.1026 &  0.0060 &  0.0041 &  0.0032 &  0.0049 & $ 7.8/ 5$ \\
161.3 & \bw &  0.0984 &  0.0051 &  0.0022 &  0.0029 &  0.0037 & $ 8.9/ 4$ \\
161.3 & \cp &  0.1012 &  0.0065 &  0.0045 &  0.0030 &  0.0042 & $ 7.7/ 4$ \\
161.3 & \ytwothree &  0.1056 &  0.0085 &  0.0019 &  0.0006 &  0.0030 & $ 1.8/ 3$ \\
\hline
172.1 & \thr &  0.1090 &  0.0077 &  0.0073 &  0.0033 &  0.0050 & $ 6.6/ 5$ \\
172.1 & \mh &  0.1036 &  0.0069 &  0.0056 &  0.0016 &  0.0035 & $ 5.8/ 4$ \\
172.1 & \bt &  0.1045 &  0.0070 &  0.0054 &  0.0016 &  0.0051 & $ 3.7/ 5$ \\
172.1 & \bw &  0.0961 &  0.0060 &  0.0037 &  0.0007 &  0.0035 & $ 2.0/ 4$ \\
172.1 & \cp &  0.1029 &  0.0074 &  0.0070 &  0.0034 &  0.0042 & $ 7.2/ 4$ \\
172.1 & \ytwothree &  0.1038 &  0.0102 &  0.0130 &  0.0012 &  0.0029 & $ 2.6/ 3$ \\
\hline
182.7 & \thr &  0.1108 &  0.0036 &  0.0026 &  0.0018 &  0.0053 & $ 0.6/ 5$ \\
182.7 & \mh &  0.1071 &  0.0032 &  0.0039 &  0.0018 &  0.0041 & $ 3.6/ 4$ \\
182.7 & \bt &  0.1110 &  0.0031 &  0.0033 &  0.0017 &  0.0059 & $ 6.0/ 5$ \\
182.7 & \bw &  0.1029 &  0.0027 &  0.0029 &  0.0004 &  0.0042 & $ 3.6/ 4$ \\
182.7 & \cp &  0.1073 &  0.0034 &  0.0027 &  0.0019 &  0.0049 & $ 0.6/ 4$ \\
182.7 & \ytwothree &  0.1096 &  0.0047 &  0.0039 &  0.0003 &  0.0034 & $ 1.3/ 3$ \\
\hline
\end{tabular}}
\end{center} 
\end{table*}

\begin{table*}[htb!]
\caption{Results of NLO+NLLA fits to event shape observable distributions
at the OPAL c.m.\ energies.
The \chisqd\ values are based on the statistical errors only.}
\label{detailedasresultsnlonlla}
\begin{center}  
\scalebox{0.9}
{\begin{tabular}{ ccrrrrrc }
\hline\noalign{\smallskip}
\roots\ [GeV] & Obs. & $\as(\roots)$ & $\pm$stat. & $\pm$exp. & $\pm$had. & $\pm$theo. & \chisqd \\
\noalign{\smallskip}\hline\noalign{\smallskip}
% was asopal_nlo+nlla_2.tex :
188.6 & \thr &  0.1143 &  0.0021 &  0.0016 &  0.0014 &  0.0058 & $14.2/ 5$ \\
188.6 & \mh &  0.1068 &  0.0019 &  0.0020 &  0.0019 &  0.0040 & $ 4.1/ 4$ \\
188.6 & \bt &  0.1128 &  0.0019 &  0.0014 &  0.0012 &  0.0061 & $ 2.7/ 5$ \\
188.6 & \bw &  0.1031 &  0.0016 &  0.0006 &  0.0006 &  0.0042 & $ 1.2/ 4$ \\
188.6 & \cp &  0.1083 &  0.0021 &  0.0013 &  0.0018 &  0.0050 & $ 4.1/ 4$ \\
188.6 & \ytwothree &  0.1075 &  0.0028 &  0.0020 &  0.0004 &  0.0032 & $ 3.4/ 3$ \\
\hline
191.6 & \thr &  0.1120 &  0.0052 &  0.0087 &  0.0016 &  0.0055 & $ 2.0/ 5$ \\
191.6 & \mh &  0.1067 &  0.0046 &  0.0055 &  0.0010 &  0.0039 & $ 5.0/ 4$ \\
191.6 & \bt &  0.1045 &  0.0051 &  0.0100 &  0.0024 &  0.0051 & $ 8.4/ 5$ \\
191.6 & \bw &  0.1002 &  0.0040 &  0.0048 &  0.0004 &  0.0039 & $ 0.9/ 4$ \\
191.6 & \cp &  0.1039 &  0.0053 &  0.0059 &  0.0020 &  0.0044 & $ 0.3/ 4$ \\
191.6 & \ytwothree &  0.1044 &  0.0070 &  0.0066 &  0.0005 &  0.0030 & $ 0.0/ 3$ \\
\hline
195.5 & \thr &  0.1130 &  0.0034 &  0.0041 &  0.0021 &  0.0057 & $ 9.3/ 5$ \\
195.5 & \mh &  0.1040 &  0.0032 &  0.0032 &  0.0021 &  0.0038 & $ 0.8/ 4$ \\
195.5 & \bt &  0.1112 &  0.0032 &  0.0028 &  0.0013 &  0.0059 & $ 4.5/ 5$ \\
195.5 & \bw &  0.1021 &  0.0027 &  0.0017 &  0.0005 &  0.0041 & $ 3.7/ 4$ \\
195.5 & \cp &  0.1073 &  0.0035 &  0.0033 &  0.0018 &  0.0049 & $ 1.7/ 4$ \\
195.5 & \ytwothree &  0.1003 &  0.0047 &  0.0046 &  0.0003 &  0.0026 & $ 1.2/ 3$ \\
\hline
199.5 & \thr &  0.1117 &  0.0036 &  0.0059 &  0.0020 &  0.0055 & $ 7.8/ 5$ \\
199.5 & \mh &  0.1045 &  0.0033 &  0.0029 &  0.0012 &  0.0039 & $ 5.2/ 4$ \\
199.5 & \bt &  0.1136 &  0.0034 &  0.0046 &  0.0022 &  0.0062 & $ 6.0/ 5$ \\
199.5 & \bw &  0.1036 &  0.0028 &  0.0019 &  0.0005 &  0.0042 & $ 2.2/ 4$ \\
199.5 & \cp &  0.1066 &  0.0036 &  0.0039 &  0.0013 &  0.0049 & $ 4.6/ 4$ \\
199.5 & \ytwothree &  0.1121 &  0.0048 &  0.0043 &  0.0009 &  0.0037 & $ 4.8/ 3$ \\
\hline
201.6 & \thr &  0.1182 &  0.0050 &  0.0054 &  0.0011 &  0.0065 & $ 1.1/ 5$ \\
201.6 & \mh &  0.1029 &  0.0048 &  0.0048 &  0.0009 &  0.0036 & $11.4/ 4$ \\
201.6 & \bt &  0.1105 &  0.0049 &  0.0034 &  0.0011 &  0.0058 & $ 6.7/ 5$ \\
201.6 & \bw &  0.1018 &  0.0040 &  0.0025 &  0.0013 &  0.0040 & $ 7.1/ 4$ \\
201.6 & \cp &  0.1091 &  0.0053 &  0.0065 &  0.0020 &  0.0053 & $ 9.7/ 4$ \\
201.6 & \ytwothree &  0.1102 &  0.0070 &  0.0073 &  0.0007 &  0.0035 & $ 1.2/ 3$ \\
\hline
204.9 & \thr &  0.1118 &  0.0032 &  0.0032 &  0.0014 &  0.0054 & $ 9.6/ 5$ \\
204.9 & \mh &  0.1109 &  0.0029 &  0.0034 &  0.0006 &  0.0045 & $ 7.9/ 4$ \\
204.9 & \bt &  0.1117 &  0.0031 &  0.0061 &  0.0011 &  0.0060 & $ 4.2/ 5$ \\
204.9 & \bw &  0.1059 &  0.0025 &  0.0031 &  0.0005 &  0.0045 & $ 5.2/ 4$ \\
204.9 & \cp &  0.1072 &  0.0032 &  0.0069 &  0.0017 &  0.0049 & $ 3.0/ 4$ \\
204.9 & \ytwothree &  0.1084 &  0.0044 &  0.0060 &  0.0006 &  0.0033 & $ 3.4/ 3$ \\
\hline
206.6 & \thr &  0.1076 &  0.0027 &  0.0019 &  0.0020 &  0.0049 & $ 2.2/ 5$ \\
206.6 & \mh &  0.1034 &  0.0025 &  0.0017 &  0.0022 &  0.0036 & $ 0.8/ 4$ \\
206.6 & \bt &  0.1126 &  0.0024 &  0.0030 &  0.0012 &  0.0061 & $ 1.2/ 5$ \\
206.6 & \bw &  0.1018 &  0.0021 &  0.0022 &  0.0005 &  0.0040 & $ 1.5/ 4$ \\
206.6 & \cp &  0.1066 &  0.0026 &  0.0020 &  0.0018 &  0.0048 & $ 1.1/ 4$ \\
206.6 & \ytwothree &  0.1052 &  0.0036 &  0.0025 &  0.0004 &  0.0030 & $ 0.5/ 3$ \\
\hline
\end{tabular}}
\end{center}  
\end{table*}

\end{appendix}

\end{document}